\documentclass[times,twocolumn,final]{elsarticle}
\usepackage[margin=1.5cm]{geometry}

\usepackage{framed,multirow}

\usepackage{amssymb}
\usepackage{latexsym}

\usepackage{url}

\usepackage{todonotes} 
\usepackage{mathtools} 
\usepackage[normalem]{ulem} 
\usepackage{svg} 
\usepackage{array} 
\newcolumntype{M}[1]{>{\centering\arraybackslash}m{#1}} 
\newcolumntype{P}[1]{>{\centering\arraybackslash}p{#1}} 
\usepackage{amsmath} 
\usepackage[linesnumbered, ruled]{algorithm2e} 

\usepackage{pdfpages} 

\usepackage{colortbl} 
\usepackage{amssymb} 
\usepackage{bm} 
\usepackage{subcaption} 
\usepackage{float} 
\usepackage{pifont}
\usepackage{graphicx} 
\newcommand{\cmark}{\ding{51}}%
\newcommand{\xmark}{\ding{55}}%
\newcommand{\blue}{\ifthenelse{10 > 100}{\textcolor{blue}}{}}
\newcommand{\red}{\ifthenelse{10 > 100}{\textcolor{red}}{}}
\newcommand{\bluetwo}{\ifthenelse{10 > 100}{\textcolor{blue}}{}}
\newcommand{\redtwo}{\ifthenelse{10 > 100}{\textcolor{red}}{}}
\usepackage{lineno}
\newcommand{\eg}{{e.g.}}
\newcommand{\ie}{{i.e.}}

\usepackage{hyperref}

\journal{Medical Image Analysis}

\begin{document}


\begin{frontmatter}

\title{DeepSMILE: Contrastive self-supervised pre-training benefits MSI and HRD classification directly from H\&E whole-slide images in colorectal and breast cancer}

\author[1,2]{Yoni Schirris}
\cortext[cor1]{Corresponding author:
  Address: Plesmanlaan 161, 1066 CX Amsterdam, the Netherlands; E-mail: j.teuwen@nki.nl
  }
\author[2,4]{Efstratios Gavves } 
\author[1]{Iris Nederlof }
\author[1]{Hugo Mark Horlings}
\author[1,2,3]{Jonas Teuwen\corref{cor1}}

\address[1]{Netherlands Cancer Institute, Plesmanlaan 121, 1066 CX Amsterdam, the Netherlands}
\address[2]{University of Amsterdam, Science Park 402, 1098 XH Amsterdam, the Netherlands}
\address[3]{Radboud University Medical Center, Department of Medical Imaging, Geert Grooteplein Zuid 10, 6525 GA Nijmegen, The Netherlands}
\address[4]{Ellogon AI B.V., The Netherlands}


\begin{abstract}
We propose a Deep learning-based weak label learning method for analyzing whole slide images (WSIs) of Hematoxylin and Eosin (H\&E) stained \blue{tumor tissue} not requiring pixel-level or tile-level annotations using Self-supervised pre-training and heterogeneity-aware deep Multiple Instance LEarning (DeepSMILE). We apply DeepSMILE to the task of Homologous recombination deficiency (HRD) and microsatellite instability (MSI) prediction.
We utilize contrastive self-supervised learning to pre-train a feature extractor on histopathology tiles of cancer tissue. Additionally, we use variability-aware deep multiple instance learning to learn the tile feature aggregation function while modeling tumor heterogeneity.
For MSI prediction in a tumor-annotated and color normalized subset of TCGA-CRC \red{(n=360 patients)}, contrastive self-supervised learning improves the tile supervision baseline from 0.77  to 0.87 AUROC, on par with our proposed DeepSMILE method.
On TCGA-BRCA \red{(n=1,041 patients)} without any manual annotations, DeepSMILE improves HRD classification performance from 0.77 to 0.81 AUROC compared to tile supervision with either a self-supervised or ImageNet pre-trained feature extractor. Our proposed methods reach the baseline performance using only 40\% of the labeled data on both datasets.
\blue{These improvements suggest we can use standard self-supervised learning techniques combined with multiple instance learning in the histopathology domain to improve genomic label classification performance with fewer labeled data.}
\end{abstract}

\begin{keyword}

\raggedright Self-supervised learning \sep Multiple Instance Learning \sep Computational pathology \sep Histogenomics
\end{keyword}

\end{frontmatter}


\section{Introduction} \label{introduction}
Early recognition of abnormalities in the DNA Damage Response (DDR) machinery in tumors can greatly support personalized medicine by identifying patients that may benefit from therapies that exploit DDR-related genomic alterations (\cite{vandervelden2019drug, pearl2015therapeutic, pilie2019state}). For example, Homologous Recombination Deficiency (HRD) and MicroSatellite Instability (MSI) can function as a biomarker indicating therapy sensitization in, respectively, breast cancer (\cite{lord-brcaness-revisited}) and  colorectal cancer (\cite{kather2018genomics, mauri2020dna, arena2020subset}) patients. 

Currently, the most common techniques for determining HRD or MSI are next-generation whole-genome and exome sequencing methods (\cite{davies-hrdetect, zhu2018novel}), targeted DNA sequencing methods using Polymerase Chain Reaction (\cite{boland1998national}), and immunohistochemistry (IHC) methods (\cite{kawakami2015microsatellite}). \blue{The former two techniques, however, are expensive, time-consuming, and not globally accessible \citep{snowsill2017molecular}. Additionally, the molecular features that indicate functional deficiencies in the homologous recombination pathway, thus indicating targeted therapy sensitivity, are inconclusive and debated  \citep{davies-hrdetect}. Therefore, these techniques are not routinely applied in the clinic. Although IHC is more affordable and is becoming common practice for, e.g., Lynch syndrome (\cite{national2017molecular-lynch-ihc}), it still requires additional laboratory testing.}

Generally, the golden standard for solid tumor diagnostics is the use of Hematoxylin \& Eosin-stained (H\&E) Whole-Slide Images (WSIs). In contrast to genome sequencing methods, H\&E WSIs are easily accessible, inexpensive, and reflect the cellular and tissue morphology that result from genomic alterations. However, the morphology of the broad range of DDR defects has not yet been described. Therefore, H\&E WSIs are, at present, not used to detect DDR defects in routine clinical diagnostics. This is a missed opportunity as employing H\&E WSIs to detect DDR defects could assist personalized medicine by guiding early patient stratification for additional diagnostic tests or guiding therapy decisions. The digitization of WSIs has opened up doors for computational analysis to perform this task. 

Recent work has shown great promise for deep learning methods for the computational analysis of digitized H\&E WSIs for genomic status classification (\cite{echle-kather-nature-clinical-msi, kather-nature-pan-cancer-histogenomics, fu-nature-pan-cancer-histogenomics, coudray-lung-histogenomics}). Although end-to-end supervised methods for a Convolutional Neural Network (CNN) on gigapixel WSIs exist, they require specialized implementations to circumvent the large memory requirements required for loss backpropagation with gigapixel images, and this leads to a low training and inference speed (\cite{pinckaers-streaming-original, chen-full-wsi-learning}). In contrast, two-stage methods stop the gradient at the tile feature extraction or tile aggregation level. Generally, these methods do not pass the entire WSI through a neural network but instead split the WSI into many small tiles which are the input to the network. Either the task is framed as tile-supervision, \ie, supervised learning from each tile to the WSI-level label after which the proportion of positively predicted tiles is said to be the WSI-level prediction (\cite{kather-nature-msi, kather-nature-pan-cancer-histogenomics, echle-kather-nature-clinical-msi, fu-nature-pan-cancer-histogenomics}), or it is framed as WSI-supervision in which the tiles are compressed using a pre-trained feature extractor to perform supervised classification of the WSI directly using all latent feature vectors of its constituting tiles as input. This can be framed as either full WSI-supervision (\cite{tellez-ext-nic-multitask, tellez-nic-lung-2021}) or weak WSI-supervision (\cite{lu-data-efficient-weakly-supervised-histo}).

However, these methods have their limitations. Since the cellular and tissue morphology related to DDR defects are unknown, there are no spatial annotations to guide training. This is why tile-supervision uses the WSI-level label as a tile-level label, which results in a noisy supervisory signal since the signal of the DDR defect is, likely, not present in every tile. Noisy supervision, in turn, leads to large data requirements (\cite{echle-kather-nature-clinical-msi}). Current WSI compression techniques report that the top-performing self-supervised learning (SSL) method is a bi-directional generative adversarial network, which is notoriously difficult to train (\cite{tellez-nic-gigapixel}) and are outperformed by supervised pre-training on large annotated datasets (\cite{tellez-ext-nic-multitask}). The field of SSL has grown quickly lately, however, and claims to close the gap between supervised and unsupervised learning in the natural image domain \citep{grill-byol, chen-simclr}. In practice, though, most deep learning methods in histopathology do not use a domain-specific feature extractor altogether by employing an ImageNet pre-trained feature extractor (\cite{coudray-lung-histogenomics, kather-nature-msi, kather-nature-pan-cancer-histogenomics, fu-nature-pan-cancer-histogenomics, lu-data-efficient-weakly-supervised-histo}). Since natural scenes and medical images have strongly different data distributions, using an ImageNet pre-trained network might not be optimal \citep{ke2021chextransfer-medical-transfer-learning, raghu2019transfusion-medical-transfer-learning}, and thus employing the latest SSL methods on unlabeled domain-specific data is a promising avenue to increase performance, generalizability, and robustness \citep{hendrycks2019using-ssl-robustness}.  \blue{Recently, work in the histopathology domain shows the value of self-supervised learning with BYOL \citep{grill-byol} and MOCOv2 \citep{chen2020improved-mocov2} when the feature extractors are frozen and treating the WSI classification problem as a multiple instance learning problem \citep{saillard2021self-owkin-dehaene, abbasi2021molecular-byol-he}}.

In this paper, we investigate the \blue{performance increase} offered by SimCLR \citep{chen-simclr}, which makes latent representations of heavily augmented versions of the same tile similar while making the latent representations of heavily augmented versions of different tiles distinct. SimCLR is used to pre-train a feature extractor without any spatial annotations or WSI-level labels. This feature extractor is evaluated and compared to the commonly used ImageNet pre-trained feature extractor on the downstream task of HRD and MSI classification. \blue{We research its relative effectiveness when finetuning the feature extractor with noisy WSI label tile-supervision in a dataset with and without tumor annotations. Additionally, we evaluate its performance increase in the case when we freeze the feature extractor to perform Attention-Based Deep Multiple Instance Learning (DeepMIL, \cite{ilse-deepmil}) on the tiles of a WSI.} Finally, we propose a feature variability-aware variant of DeepMIL, which models tumor heterogeneity and further increases performance.

\subsection{Contributions}
\red{The main contributions of this work can be summarized as follows: 1) We outperform existing tile-supervised WSI-label learning methods for MSI classification for a tumor-annotated and color-normalized colorectal cancer dataset using a feature extractor pre-trained with SimCLR. 2) We show that this improvement is not seen when predicting HRD in a dataset without tumor annotations, and propose a method to improve this performance. 3) We propose a Deep learning-based weak label learning method for histopathology not requiring pixel-level or tile-level annotations using Self-supervised pre-training and heterogeneity-aware deep Multiple Instance LEarning (DeepSMILE). DeepSMILE compresses a whole-slide image with a self-supervised pre-trained feature extractor and uses VarMIL, which extends DeepMIL with a feature variability module to model tumor heterogeneity. DeepSMILE outperforms the tile supervision baselines and DeepMIL for HRD classification, and focuses on features that are expected to be related to DDR deficiencies.}
\\

The paper is structured as follows: Section~\ref{section-intro-related-work} provides an overview of previous work and positions our work within it. Section~\ref{section-mm} describes the data used, followed by section~\ref{section-model} describing our model details. Section~\ref{experiments} describes all experiments. We conclude with the discussion and conclusion in section~\ref{discussion}.

\section{Related work}
\label{section-intro-related-work}
We distinguish between two-stage (not end-to-end) and one-stage (end-to-end) methods. DeepSMILE falls in the former category, and we position our work in relation to other two-stage methods in Table \ref{table-related-work}.

Two-stage methods generally cut the WSI into many small cropped patches, the \textit{tiles}, which are processed to classify the WSI. In two-stage methods, one can distinguish between tile-supervised WSI-label methods, fully supervised, and weakly supervised methods.

\subsection{Two-stage: tile-supervised WSI-label methods}
Two-stage tile-supervised WSI-label methods assign the WSI-level label to each tile of a region of interest of a WSI and treat the tile classification as a supervised learning task. Most commonly, the patient-level label is computed as the fraction of positively classified tiles. This method has been applied to gene mutation classification in lung cancer (\cite{coudray-lung-histogenomics}), MSI status classification in colorectal and stomach cancer (\cite{kather-nature-msi, echle-kather-nature-clinical-msi}), and a large variety of genomic, transcriptomic, and survival labels in pan-cancer tissue (\cite{kather-nature-pan-cancer-histogenomics, fu-nature-pan-cancer-histogenomics, echle-kather-nature-clinical-msi}). 

\subsection{Two-stage: fully supervised methods}
Two-stage fully supervised methods reduce the dimensionality of the WSI by replacing each tile in place with its latent feature representation as encoded by a pre-trained feature extractor. The resulting compressed WSIs are subsequently used as input for fully supervised methods. Several methods can be employed to pre-train the feature extractor. Self-supervised methods \citep{tellez-nic-gigapixel} and multi-task supervised pre-training on other tasks and data  \citep{tellez-ext-nic-multitask, tellez-nic-lung-2021} have been described for this task in the literature.

\subsection{Two-stage: weakly supervised methods}
In contrast to fully supervised methods, two-stage weakly supervised methods use a MIL approach. In a MIL approach, it is assumed that some of the unlabeled tiles (or \textit{instances}) in the labeled WSI (or \textit{bag}) contain the signal for the WSI-level label. These methods propose a linear or non-linear combination of the latent features or predicted scores of a selection of the tiles in a WSI to represent the WSI latent features. This WSI latent feature is then used to classify the WSI, so that a WSI-level loss can be computed and backpropagated through the classification network.

\cite{durand-weldon} compute a score for each instance and use the top and bottom $R$ scores to compute the bag-level prediction. \cite{courtiol-nature-mesothelioma} extend the work by \cite{durand-weldon} to tissue segmentation and WSI-level label classification using an MLP on the top and bottom instance scores. As in our work, \cite{ilse-deepmil} learn the bag-level classification by computing an attention-weighted mean of all tile feature vectors. This latent representation of the WSI is fed into an MLP to compute the bag label. \cite{lu-cpc-deepmil} used DeepMIL and showed that pre-training a feature extractor with contrastive predictive coding (\cite{oord-cpc}) on histopathology tiles improves downstream malignancy classification performance when compared to an ImageNet pre-trained feature extractor, yet this has not been evaluated on full WSIs. Additionally, \cite{campanella-nature-clinical-weak} produce tile-level feature vectors and class scores using a network trained with a max-pooling MIL approach. The top 20 scored feature vectors are passed to an RNN to produce the WSI-level label to detect metastasis. Similarly, \cite{valieris2020deep-mmrd-hrd} have applied this method to predict HRD and mismatch repair deficiency, the DDR deficiency that is closely related to MSI. \red{Recently, self-supervised pre-training has been shown to be effective for MSI prediction using MoCo V2 \citep{saillard2021self-owkin-dehaene} and breast cancer subtype classification using BYOL \citep{abbasi2021molecular-byol-he} in a MIL setting, and various more involved MIL classifiers have been developed that, compared to vanilla DeepMIL, alter the attention computation or ranking (\cite{lu-data-efficient-weakly-supervised-histo, bilal2021development-msi-ranking}).}

\begin{table*}[!t]
\footnotesize
\caption{\label{table-related-work}Overview of two-stage tile-supervised and weak label learning methods for WSI-level label classification.}
\centering
    \begin{tabular}{ m{1.8cm} m{3.6cm} M{1.6cm} M{1.6cm} M{1.6cm} M{1.6cm} M{1.6cm} M{1.6cm}}
    \hline
        \raggedright{Class of methods}  & Method(s) & \raggedright{Domain-specific feature extractor} & \raggedright{Evaluates MIL} & \raggedright{Evaluates tile supervision} & \raggedright{Applied to WSIs} & \raggedright{Models feature variability} & 
        \raggedright \arraybackslash Predicts genomic labels \\ \hline
        
        \raggedright{Two-stage tile-supervised WSI label learning} & 
        \cite{kather-nature-msi, kather-nature-pan-cancer-histogenomics}\newline
        \cite{echle-kather-nature-clinical-msi}\newline
        \cite{coudray-lung-histogenomics}\newline
        \cite{fu-nature-pan-cancer-histogenomics}\newline
        \cite{muti2021development} & 
        \cellcolor{orange!25}\cmark / \xmark  &
        \cellcolor{red!25}\xmark & 
        \cellcolor{green!25}\cmark &
        \cellcolor{green!25}\cmark & 
        \cellcolor{red!25}\xmark & 
        \cellcolor{green!25}\cmark \\ \hline
        
        \multirow{10}{2.3cm}{Two-stage weak label learning} 
        & \raggedright{\cite{valieris2020deep-mmrd-hrd}} &
         \cellcolor{red!25}\xmark & 
         \cellcolor{green!25}\cmark & 
         \cellcolor{red!25}\xmark & 
         \cellcolor{green!25}\cmark & 
         \cellcolor{orange!25}\cmark / \xmark &
         \cellcolor{green!25}MSI, HRD 
         \\
        
        & \raggedright{\cite{campanella-nature-clinical-weak}} &
         \cellcolor{red!25}\xmark & 
         \cellcolor{green!25}\cmark & 
         \cellcolor{red!25}\xmark & 
         \cellcolor{green!25}\cmark & 
         \cellcolor{orange!25}\cmark / \xmark &
         \cellcolor{red!25}\xmark 
         \\

        & \raggedright \cite{courtiol-nature-mesothelioma} &
        \cellcolor{red!25}\xmark & 
        \cellcolor{green!25}\cmark & 
        \cellcolor{red!25}\xmark & 
        \cellcolor{green!25}\cmark &    
        \cellcolor{red!25}\xmark & 
        \cellcolor{red!25}\xmark \\

          & \cite{lu-predict-origin-of-cancer, lu-data-efficient-weakly-supervised-histo} & 
         \cellcolor{red!25}\xmark & 
         \cellcolor{green!25}\cmark & 
         \cellcolor{red!25}\xmark & 
         \cellcolor{green!25}\cmark & 
         \cellcolor{red!25}\xmark & 
         \cellcolor{red!25}\xmark 
         \\

         & \cite{ilse-deepmil} & 
         \cellcolor{red!25}\xmark & 
         \cellcolor{green!25}\cmark & 
         \cellcolor{red!25}\xmark & 
         \cellcolor{red!25}\xmark & 
         \cellcolor{red!25}\xmark & 
         \cellcolor{red!25}\xmark
         \\
         
         & \cite{lu-cpc-deepmil} & 
         \cellcolor{green!25}CPC &
         \cellcolor{green!25}\cmark & 
         \cellcolor{red!25}\xmark & 
         \cellcolor{red!25}\xmark &
         \cellcolor{red!25}\xmark & 
         \cellcolor{red!25}\xmark       \\
         
          & \cite{bilal2021development-msi-ranking} & 
         \cellcolor{red!25}\xmark & 
         \cellcolor{green!25}\cmark & 
         \cellcolor{red!25}\xmark & 
         \cellcolor{green!25}\cmark & 
         \cellcolor{red!25}\xmark & 
         \cellcolor{green!25}MSI \\
         
         & \cite{saillard2021self-owkin-dehaene} & 
         \cellcolor{green!25}MoCov2 & 
         \cellcolor{green!25}\cmark & 
         \cellcolor{red!25}\xmark & 
         \cellcolor{green!25}\cmark & 
         \cellcolor{red!25}\xmark & 
         \cellcolor{green!25}MSI \\
         
         & \cite{abbasi2021molecular-byol-he} & 
         \cellcolor{green!25}BYOL & 
         \cellcolor{green!25}\cmark & 
         \cellcolor{red!25}\xmark & 
         \cellcolor{green!25}\cmark & 
         \cellcolor{red!25}\xmark & 
         \cellcolor{orange!25} molec. subtype \\
        
         & \textbf{DeepSMILE (Ours)} & 
         \cellcolor{green!25}SimCLR & 
         \cellcolor{green!25}\cmark &
         \cellcolor{green!25}\cmark & 
         \cellcolor{green!25}\cmark & 
         \cellcolor{green!25}\cmark & 
         \cellcolor{green!25}MSI, HRD\\
    \end{tabular}
\end{table*}

\subsection{End-to-end methods: contextless}
\cite{xie-end-to-end-part-learning} cluster the latent features of tiles of a WSI into representative \textit{parts} and concatenate each part's representative tile. This concatenation is mapped to the WSI-level label by a fully connected layer. For prostate cancer classification, they obtain similar results to the MIL-RNN method by \cite{campanella-nature-clinical-weak}. Although training is end-to-end, this method only concatenates tile-level features and thus does not model tile interactions to include higher-level spatial context.

\subsection{End-to-end methods: context-aware}
\cite{pinckaers-streaming-original} reduce the memory requirements of a CNN by 97\% with a streaming CNN, which allows applying the CNN directly to megapixel histopathology images. \cite{pinckaers-streaming-prostate-cancer} further extended the method to accommodate gigapixel WSIs, applied to prostate cancer detection, reaching similar results to the MIL-RNN method by \cite{campanella-nature-clinical-weak}. \cite{chen-full-wsi-learning} instead leverage the unified memory mechanism and other GPU optimization techniques to overcome the memory constraints. Although these methods reach state-of-the-art results, streaming CNNs are approximately $16\times$ slower than MIL methods during training and inference on WSIs at low resolution, which scales non-linearly with increasing resolution.
\\

Our proposed method, DeepSMILE, is a two-stage weakly supervised method applied to HRD and MSI classification. \blue{Compared to the existing methods, it introduces and evaluates the use of SimCLR, a contrastive self-supervised learning method, to pre-train a histopathology-specific feature extractor used for tile supervision models and multiple instance learning models.} Furthermore, we introduce an extended version of DeepMIL that models tumor heterogeneity. We describe our model in detail in section~\ref{section-model}.

\section{Materials and methods} \label{section-mm} 

\subsection{Data} \label{section-mm-data} 
We use digitized H\&E tissue slides of breast (BC) and colorectal (CRC) cancer tissue to classify a tumor's genomic labels. These WSIs are large gigapixel images that can exceed 100,000$\times$100,000 px, whereas the genomic labels are binary representations of complex genomic features derived from DNA sequencing results. This section describes the data collection and pre-processing steps for each dataset, and how the genomic labels are obtained. \red{A detailed description of demographic variables (age, race, AJCC pathologic stage, prior treatment, gender, genomic label) per data split can be found in the Supplementary materials.}

\subsubsection{Colorectal cancer tissue dataset: TCGA-CRCk} \label{data-crck}
We use the pre-processed \red{(tiled from tumor annotations and color-normalized)} colorectal tumor tiles extracted from Formalin-Fixed, Paraffin-Embedded (FFPE) WSIs from the Cancer Genome Atlas (TCGA) with accompanying binarized MSI labels from \cite{kather-msi-dataset}. We use the train-test split as given and perform a 5-fold patient-level train-validation split.
This dataset consists of 192,314 (train: 93,408, test: 98,906) tiles from 360 patients (train: 260, test: 100). The train and test set respectively consist of 39 and 26 MSI, and 221 and 74 MicroSatellite Stable (MSS) patients. \blue{These patients are from multiple centers in the United States of America}. We refer to this dataset as \textit{TCGA-CRCk}.

\subsubsection{Breast cancer tissue dataset: TCGA-BC} \label{data-tcgabc-basis}
We obtain all FFPE WSIs for BC tissue from TCGA (referred to as \textit{TCGA-BC}), \blue{which is collected from multiple centers in the United States of America}, and retrieve genomic DDR-related labels, including HRD Score, from \cite{knijnenburg-molecular-landscape-ddr-tcga}. The HRD Score (\cite{hrdscore-1-timms2014association, hrdscore-2-marquard2015pan}) is a discrete score computed as the sum of the number of subchromosomal regions with allelic imbalance extending to the telomere (\cite{hrdscore-tai-birkbak2012telomeric}), the number of chromosomal breaks between adjacent regions of at least 10 megabases (\cite{hrdscore-lst-popova2012ploidy}), and the number of regions with a loss of heterozygosity event of intermediate size (\cite{hrdscore-hrdloh-abkevich2012patterns}). Since the relationship of the HRD Score to actual homologous recombination functionality and cellular morphology is not known, we test two binarization strategies to define the classification task of distinguishing HRD-high (assumed to be HR deficient and thus sensitive to targeted therapy) and HRD-low patients (assumed to be HR proficient and thus not sensitive to targeted therapy). First, we split the score at the median (mHRD), similar to previous work \citep{kather-nature-pan-cancer-histogenomics}. Secondly, we aim to provide a better supervisory signal by splitting the set in tertiles (tHRD) and assigning a top-tertile or bottom-tertile label, discarding patients with an HRD score close to the median. That is,
\begin{equation} \label{eq-median-split}
    \text{mHRD}^m = 
    \begin{cases}
    1, & \text{if } \text{HRD}^m > q(0.5; D^\text{TCGA-BC}) = 21\\
    0, & \text{otherwise}
    \end{cases}
\end{equation}

where $\text{HRD}^m$ is the HRD score of patient $m$ and $q(\alpha; D^d)$ indicates the $\alpha$-quantile of the HRD scores of dataset $d$. For tHRD:
\begin{equation} \label{eq-tertile-split}
    \text{tHRD}^m = 
    \begin{cases}
    1, & \text{if } \text{HRD}^m \geq q(0.66; D^{\text{TCGA-BC}}) = 13\\
    0, & \text{if } \text{HRD}^m \leq q(0.33; D^{\text{TCGA-BC}}) = 30\\
    & \text{discard patient otherwise}
    \end{cases}
\end{equation}

The train-test splits for TCGA-BC are generated for those patients provided by \cite{knijnenburg-molecular-landscape-ddr-tcga} for which HRD labels and FFPE WSIs are available. The splits are patient-level \red{stratified for mHRD score. We create 5-fold cross-validation splits with 20\% (215-231 patients) in the test set, 20\% (219-229 patients) in the validation set, and 60\% (660-670 patients) in the training set. The exact patient distributions and their demographic and clinical variables can be found in the Supplementary materials.}

\red{The background of WSIs is segmented using the FESI (\cite{bug2015foreground-fesi}) algorithm on the WSI at 10 microns per pixel, and tiles of which more than half of the pixels is covered by the background segmentation mask are discarded. The remaining tissue tiles are then tiled at a spacing of 1.14 microns per pixel (mpp) with a tile size of $224 \times 224$px, resulting in an edge length of 256 µm}. We selected the zoom level best representing this spacing, before downsampling. \red{We use all extracted tissue tiles for all training and inference pipelines, resulting in 2,889,870 tiles for 1,127 WSIs of 1,041 patients.}

\subsection{Evaluation} \label{section-evaluation}
The model performance is evaluated using the area under the receiving operator characteristic curve (AUROC, or AUC) \blue{and F1 score at a cut-off of 0.5 on slide level predictions} compared to the genomic binary labels. For each experiment, we report the mean and standard deviation of the AUC \blue{and F1} scores on the test split for 5 different folds and visualize the ROC \blue{and precision-recall} curves. \red{To statistically compare two methods we use the bootstrapping method. More specifically, for each fold we sample 40 random samples with replacement from the test set 100 times and compute the metric for each bootstrapped sample. We then use a two-sided student's t-test to compare mean of the distribution of bootstrapped metric values, assuming independence between samples and similar variance. With a $p<0.05$ we reject the null hypothesis that the metrics are similarly distributed, and conclude that the model with a higher mean performs significantly better.} The complete experimental pipeline is summarized in Algorithm \ref{mm-steps}. We use common training, intermediate validation, and final validation steps as presented in Algorithm \ref{mm-steps} line \ref{mm-steps-start-fold-training}-\ref{mm-steps-end-fold-training}. We select the model with the highest patient-level AUC on the validation set for final validation on the test set.

\begin{algorithm} 
Download WSIs with matching genomic labels \hfill[sec \ref{section-mm-data}] \label{mm-steps-obtain-data}\\

Extract tissue-containing tiles from WSI \hfill[sec \ref{data-tcgabc-basis}] \label{mm-steps-pre-processing-data} \\

Split tiles on patient level into 5-fold cross-validation train, val (and test) split \hfill[sec \ref{data-tcgabc-basis}] \label{mm-steps-split-data}\\

Perform self-supervised pre-training on tiles from train set, save model weights, extract features from training and test tiles and save to disk \hfill[sec \ref{section-model-ssl}] \label{mm-steps-ssl}\\ 
\uIf{\upshape \texttt{pipeline == "baseline"}}{\texttt{model} = ImageNet pre-trained CNN with tile-level MLP classifier \hfill[sec \ref{section-model-baseline}] \label{mm-steps-tile-supervised}\\
\texttt{data\_loader} = load tiles as samples, with patient-level label per tile}
\ElseIf{\upshape \texttt{pipeline == "DeepSMILE"}}
{\texttt{model} = VarMIL WSI-level classifier \hfill[sec \ref{section-model-varmil}] \label{mm-steps-deepmil}\\
\texttt{data\_loader} = load latent features of all tiles of a WSI as samples, with patient-level label per sample }
\texttt{optimizer = ADAM} \hfill[sec \ref{section-model}] \\
\For{\upshape \texttt{fold in k\_folds}}{ \label{mm-steps-start-fold-training}
    
        \For{\upshape \texttt{(x, y, step) in data\_loader}}{
            \texttt{pred = model.forward(x)}\\
            \texttt{loss = \blue{CE}(y, pred)} \hfill[sec \ref{section-model}] \label{mm-steps-loss}\\
            \texttt{loss.backward()}\\
            \texttt{optimizer.step()}\\
            \If{\upshape \texttt{step \% evaluate\_every == 0}  \label{mm-steps-evaluation}} {
                \texttt{val\_pred = \texttt{model.forward(X\_val)} \\
                AUC = compute\_auc(val\_pred,  Y\_val)} \hfill [sec \ref{section-evaluation}] \\
                \texttt{save(AUC, model)}\\
            }
        }
    
    Pick model with top AUC on val set \hfill[sec \ref{section-evaluation}] \label{mm-steps-model-picking}\\
    Use best model to evaluate performance on test set \label{mm-steps-end-fold-training}
}
Report k-fold $\mu\pm\sigma$ of \blue{slide-level AUC and F1} \hfill[sec \ref{experiments}] \label{mm-steps-metrics}\\
\caption{\label{mm-steps} Summary of experimental pipeline. Square brackets contain section number providing details for that step.}
\end{algorithm}


\section{Model} \label{section-model}
\begin{figure*}[!t] 
\centering
\includegraphics[width=\hsize]{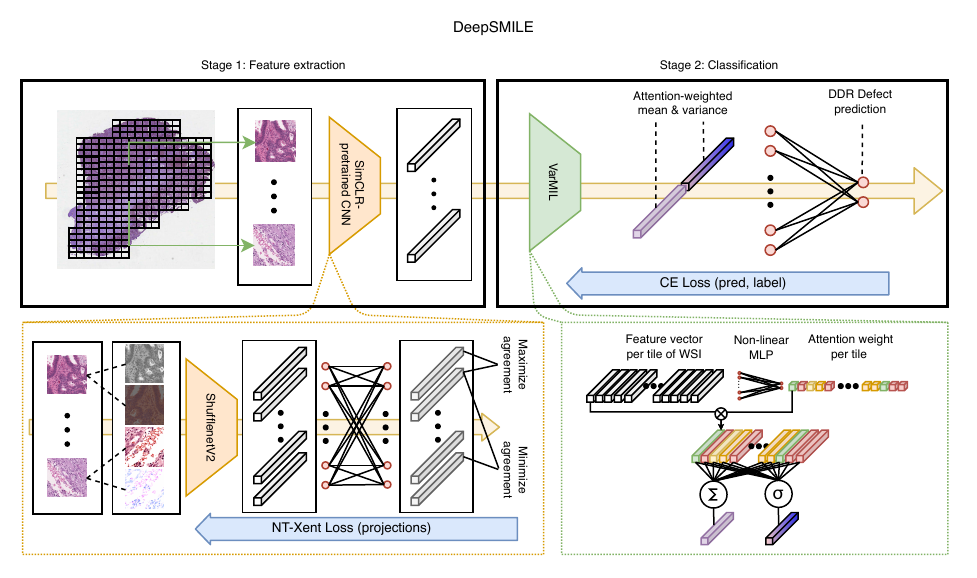}
\caption{A visual summary of DeepSMILE. Top-left: We extract all non-background tiles from a WSI, and from each tile extract the latent feature vector using a domain-specific feature extractor pre-trained with SimCLR. 
Bottom-left: During pre-training, SimCLR transforms each image twice with random augmentations, extracts latent features using the backbone of interest, maps each vector to a projection head, and uses these to compute the loss. After training, the projection MLP is discarded and the backbone of interest is used to compress tiles. 
Top-right: VarMIL computes the attention-weighted mean and variance of all tile latent feature vectors. The concatenated mean and variance are passed to a linear classification layer to compute the WSI-level genomic label classification, and the loss is backpropagated through the right side of the system.
Bottom-right: VarMIL computes the attention weight from each latent feature vector, and computes an attention-weighted mean and variance of all tile latent feature vectors of the WSI. }
\label{fig-model-as-a-system}
\end{figure*}

In this section we describe our model, which is a two-stage weakly supervised method, using self-supervised pre-training and feature variability-aware deep multiple instance learning. We will refer to this model as DeepSMILE (from Self-supervised heterogeneity-aware Multiple Instance LEarning). Our model is visualized in Figure~\ref{fig-model-as-a-system} and its comparison to existing methods is summarized in Table \ref{table-related-work}.

We split this section into two parts. In the first part in section~\ref{section-model-ssl}, we present how we train an in-domain pathology-specific feature extractor with SimCLR to extract the latent feature vector of each tile. In the subsequent section~\ref{section-model-varmil}, we present VarMIL, which comprises the second stage of the model. This is a MIL approach extending DeepMIL, detailed in section~\ref{section-model-deepmil}, that models the variance of each latent feature of all instances in a bag.

In the experiments (section~\ref{experiments}), we will compare DeepSMILE to a tile-supervised WSI-label learning baseline with an ImageNet pre-trained feature extractor as presented in section~\ref{section-model-baseline}.

\subsection{Feature extraction} \label{section-model-extraction}
\subsubsection{Self-supervised pre-training} \label{section-model-ssl}
In the feature extraction stage, we use a CNN to map each tile of a WSI to a lower-dimensional representation. Formally, this neural network $z_\theta$ parametrized by $\theta$ maps the $i^\text{th}$ tile of patient $m$, $X_i^m$, into an $H$-dimensional vector $Z_i^m$: 
\begin{equation} \label{eq-ssl-zim}
Z_i^m := z_\theta(X_i^m)
\end{equation}
To find the parameters $\theta$ we use SimCLR, which is a contrastive self-supervised learning method. We thus obtain an in-domain histopathology-specific feature extractor without requiring pixel-level annotations or WSI-level labels. This is particularly relevant when the cellular and tissue morphologies related to the genomic label are unknown, as is the case with HRD and MSI.

The feature extractor $z_\theta$ is initialized using He initialization \citep{he-rectifiers}. The loss function, data augmentations and their parameters are identical to those in \cite{chen-simclr}; we use the normalized temperature-scaled cross-entropy (NT-Xent) loss on projected latent feature vectors of images that are augmented using a random crop, random horizontal flip, color jitter, and a random grayscale transformation. A visual representation of SimCLR is provided in the lower-left frame of Figure~\ref{fig-model-as-a-system}. The parameters $\theta$ are optimized using the Adam optimizer with a learning rate of $3 \times 10^{-4}$, $\beta_1=0.9$, $\beta_2=0.99$, with no learning rate schedule or weight decay.

Similar to previous work \citep{kather-nature-msi, kather-nature-pan-cancer-histogenomics, echle-kather-nature-clinical-msi} we use different networks for TCGA-CRCk and TCGA-BC, where we use Resnet18 \citep{he-residual-learning} and ShufflenetV2 \citep{ma-shufflenet} respectively. This design choice was made to be able to directly compare the results. This difference in networks and dataset size leads to different choices in batch size and number of epochs. On TCGA-BC we train $z_\theta$ for \red{60 epochs with a batch size of $1024$}, whereas we use 100 epochs and a batch size of $128$ for TCGA-CRCk. The pre-training respectively takes approximately three days using four or one Nvidia Titan RTX card(s) (24GB VRAM). \red{The code and run configuration can be found on \url{https://github.com/NKI-AI/hissl}.}

\subsection{Classification} \label{section-model-classification} \label{section-mm-loss} 
To predict the genomic label, which is a whole-slide level label, we introduce VarMIL, an extension of DeepMIL which additionally takes the intra-WSI inter-tile variance of extracted tile features into account.
We compare with two baseline models: DeepMIL and a tile-supervision baseline with a tile-level majority vote.
For each classification method, we use the \blue{cross-entropy loss.} \red{The code and run configuration can be found at \url{https://github.com/NKI-AI/dlup-lightning-mil}.}


As our method extends DeepMIL, we first introduce the baseline models in sections~\ref{section-model-baseline} and \ref{section-model-deepmil}.

\subsubsection{Baseline: tile supervised WSI-label learning} \label{section-model-baseline}
We compare against a tile-supervised method as used in, \eg, \cite{kather-nature-msi, kather-nature-pan-cancer-histogenomics} and \cite{fu-nature-pan-cancer-histogenomics}. In these models, the patient-level label is assigned to each tile from the tumor bed, and the classification task is defined to predict the WSI-level label directly from a single tile. The final WSI-level prediction is defined as a variant of the majority vote, computed as \red{the mean of all tile predictions, that is:
\begin{equation}
    p^m=\frac{1}{N} \sum_{i=1}^{I_m} p^m_i,
\end{equation}
where $I_m$ is the number of tiles for patient $m$ and $p_i^m$ is the class probability of tile $i$ for patient $m$ belonging to the positive class}.

\red{On TCGA-BC we follow the approach of \cite{kather-nature-pan-cancer-histogenomics} and train a ShufflenetV2 for 10,000 steps (approximately $4$ epochs) with a learning rate of $5\times10^{-5}$ and a batch size of $512$. Whereas, on TCGA-CRCk, we follow the approach of \cite{kather-nature-msi} and train a ResNet18 for 20,000 steps (approximately 100 epochs) with a batch size of $256$, learning rate of $10^{-6}$ (selected following a hyperparameter search in the original paper). For both datasets, we use the Adam optimizer with $\beta_1=0.9$, $\beta_2=0.99$, without a learning rate schedule, and evaluate the performance every 100 steps. Each image is augmented with a random flip-transform (horizontal flip with $p=0.5$, vertical flip with $p=0.5$, random rotation of $0$, $90$, $180$, or $270$ degrees with equal probability.}

The convolutional layers are initialized using either an ImageNet pre-trained network or a self-supervised pre-trained network on histopathology tiles as presented in section~\ref{section-model-ssl}. The final fully connected layers are re-initialized using He initialization. In contrast to \cite{kather-nature-pan-cancer-histogenomics} we fine-tune all layers instead of only the last layers.

\subsubsection{Weak label classification: DeepMIL}\label{section-model-deepmil}
DeepMIL is a permutation-invariant MIL model that represents the bag-level latent features as the attention-weighted average of instance-level feature vectors. This bag-level representation is subsequently classified by a fully connected layer. DeepMIL is applied to the encoded tile feature vectors from the first stage as described in section~\ref{section-model-extraction}.

The DeepMIL algorithm predicts for each patient $m$ an attention weight vector $a^m := (a_1^m, a_2^m, \ldots, a_{I_m}^m)$ where $I_m$ are the number of available tiles for this patient. This is computed as the softmax of the output of a two-layer multilayer perceptron (MLP) with weights $(W_i, b_i)_{i=1,2}$ on top of all tile-level feature vectors $Z^m$:
\begin{equation}
a^m := \operatorname{softmax}\left(W_2 \tanh \left(W_1 Z^m + b_1 \textbf{1}_{1 \times I_m}\right) + b_2\right),
\end{equation}
where the dimensionalities of the weights $(W_i, b_i)_{i=1,2}$ depend on the output dimension $H$ of the feature extractor of the first stage and the MLP dimension $\nu$. In particular $W_1 \in \mathbb{R}^{\nu \times H}$, $b_1 \in \mathbb{R}^{\nu}$ and $W_2 \in \mathbb{R}^{1 \times \nu}$ and $b_2 \in \mathbb R$. We define $\textbf{1}_{1\times I_m}$ as $\left[1 \cdots 1\right] \in \mathbb{R}^{1 \times I_m}$, meaning that $b_1$ is added tile-wise to each $\nu$-dimensional column of $W_1Z^m$. The output of the MLP is $1$-dimensional for each tile, and therefore $a^m \in \mathbb{R}^{I_m}$.

Subsequently the WSI-level representation $\bar{Z}^m \in \mathbb{R}^{H}$ is computed as the matrix-vector product between the attention vector $a^m$ and the tile feature vector matrix $Z^m \in \mathbb{R}^{H \times I_m}$ from \eqref{eq-ssl-zim}:
\begin{equation} \label{eq-attention-weighted-mean}
    \bar{Z}^m := Z^m a^m = \sum_{i=1}^{I_m} a_i^m Z_i^m,
\end{equation} 

Finally, the two class output probabilities $p^m = (p_1^m, p_2^m)$ are computed by a linear layer with trainable weights $(W_\xi, b_\xi)$ on top of the WSI-level latent representation $\bar{Z}^m$:
\begin{equation}
    p^m = \text{sigmoid} \left( W_\xi \bar{Z}^m + b_\xi \right),
\end{equation}
where $W_\xi \in \mathbb{R}^{2 \times H}$ and $b \in \mathbb{R}^2$.
As in \cite{ilse-deepmil} we select $\nu = 128$ and use the Adam optimizer with $\beta_1=0.9$ and $\beta_2=0.99$. The parameters are initialized using the He initialization. \red{The learning rate and weight decay were found using grid search}. For the learning rate and weight decay, this results in values of $5 \times 10^{-4}$ and $10^{-4}$, respectively. \red{On both TCGA-BC and TCGA-CRCk we train for 20,000 steps with a batch size of 1 and evaluate the performance every 20 steps.}

\subsubsection{Our method: Weak label classification with VarMIL} \label{section-model-varmil}
A limitation of DeepMIL is that the attention-weighted mean is unable to capture tile interactions and global, high-level features. This results in an aggregated feature vector that only represents local, tile-level features. However, global features, such as tumor border shape and intratumor heterogeneity, might be indicative of HRD or MSI. Therefore, we extend DeepMIL with an attention-weighted variance module, termed \textit{VarMIL}, which computes the variability in features across tiles within a single WSI as a measure of tissue heterogeneity. 
In addition to $\bar{Z}^m$ from the original DeepMIL framework, we propose to model the feature variability of the tiles by adding a learned attention-weighted variance. The weighted variance for any patient $m$ is defined as 

\begin{equation} \label{eq-attention-weighted-std}
Z_\sigma^m
= 
    \frac{I_m}{I_m-1}
    \sum_{i=1}^{I_m}a_i^m(Z_i^m - \bar{Z}^m)^2
\end{equation}
We then concatenate $Z_\sigma^m$ and  $\bar{Z}^m$ into a single vector, that is 
\begin{equation}
    \hat{Z}^m \coloneqq \left[ \begin{array}{c}  \bar{Z}^m \\ Z_\sigma^m \end{array} \right],
\end{equation}
which is the novel WSI-level latent representation that is passed to a linear layer with trainable weights ($W_\psi, b_\psi$):
\begin{equation}
p^m= \text{sigmoid} \left( W_\psi \hat{Z}^m + b_\psi \right),
\end{equation}
where $W_\psi \in \mathbb{R}^{2\times 2H}$ and $b_\psi \in \mathbb{R}^2$.

We use the same hyperparameters and initialization as used for DeepMIL (section~\ref{section-model-deepmil}).

    \begin{table*}[!t]
    \small
    \caption{\label{results-table-auc}
    Comparison of DeepSMILE (SimCLR-VarMIL) to the baseline (ImageNet pre-trained tile-supervised WSI-label learning) using the area under the receiver operating characteristic curve. Top row shows the results of our baseline model, second row shows same method but with a SimCLR pre-trained network. Third (fifth) row shows weak label learning with DeepMIL (VarMIL) on features extracted with an ImageNet pre-trained feature extractor, while the fourth (sixth) row uses a SimCLR pre-trained feature extractor. The top row of the rightmost column reproduces the same results of \cite{kather-nature-msi} (reported as 77 (95\% CI, 62–87)) with the same method on the same dataset. The following rows show the added effect of each our our proposed modules, illustrating that self-supervised learning improves performance for both tile supervision and multiple instance learning methods. \textbf{Bold} indicates statistically significant ($p<0.05$) greater value than second greatest score.}
    \centering 
    \begin{tabular}{p{2.5cm} p{2.5cm} p{1.5cm} p{1.5cm} p{1.5cm} p{1.5cm} p{1.5cm} p{1.5cm}}
    \hline
         \multirow{2}{2.5cm}{Extractor initialization}   &  \multirow{2}{2.5cm}{Classification Method} &  \multicolumn{4}{c}{TCGA-BC} & \multicolumn{2}{c}{TCGA-CRCk} \\ 
         &  & \multicolumn{2}{c}{mHRD} & \multicolumn{2}{c}{tHRD} & \multicolumn{2}{c}{MSI} \\
         & & \multicolumn{1}{c}{AUROC} & \multicolumn{1}{c}{F1} & \multicolumn{1}{c}{AUROC} & \multicolumn{1}{c}{F1} & \multicolumn{1}{c}{AUROC} & \multicolumn{1}{c}{F1} \\
         \hline
         
         ImageNet & Tile supervision  & $0.71\pm0.03$ & $0.64\pm0.04$ & $0.78\pm0.03$ & $0.71\pm0.04$  & $0.77\pm0.06$ &  $0.46\pm0.18$ \\
         SimCLR & Tile supervision    & $0.71\pm0.04$ & $0.63\pm0.05$ & $0.78\pm0.06$ & $0.72\pm0.04$ & \bm{$0.87\pm0.01$} &  \bm{$0.61\pm0.11$} \vspace{0.3cm} \\ 
           
         ImageNet & DeepMIL         & $0.68\pm0.02$  & $0.49\pm0.27$ & $0.69\pm0.04$ & $0.69\pm0.03$  & $0.65\pm0.01$ & $0.26\pm0.14$ \\ 
         SimCLR & DeepMIL           & $0.75\pm0.04$  & \bm{$0.68\pm0.03$} & \bm{$0.79\pm0.05$} & \bm{$0.71\pm0.01$}  & $0.85\pm0.03$ & $0.36\pm0.23$ \vspace{0.3cm} \\
          
         ImageNet & VarMIL          & $0.68\pm0.03$  & $0.44\pm0.31$ & $0.71\pm0.05$ & $0.68\pm0.03$ & $0.69\pm0.04$ & $0.43\pm0.04$  \\
         SimCLR & VarMIL            & \bm{$0.75\pm0.03$}  & \bm{$0.69\pm0.04$} & \bm{$0.81\pm0.04$} & \bm{$0.72\pm0.05$} & $0.86\pm0.02$ & $0.47\pm0.10$ \\
    \end{tabular}
    \end{table*}

\section{Experiments} \label{experiments}

\subsection{Experiment 1: Self-supervised learning improves MSI classification on TCGA-CRCk for MIL and tile supervision}
\red{
We perform an ablation study to show the added value of each of our proposed components on the downstream task of MSI classification on the TCGA-CRCk dataset (for dataset details, see section~\ref{data-crck}). Since this is a published pre-processed dataset, we can directly compare the results with those from \cite{kather-nature-msi}, \cite{saillard2021self-owkin-dehaene} and \cite{bilal2021development-msi-ranking}.
From Table \ref{results-table-auc} we see that our ImageNet-initialized tile-supervision baseline achieves an AUC of $0.77\pm0.06$, similar to the published results in \cite{kather-nature-msi}. In row 2 one can see that a SimCLR pre-trained feature extractor increases AUC by $0.1$ up to $0.87\pm0.01$, and increases F1 score from $0.46\pm0.18$ up to $0.61\pm0.11$ for the tile-supervision method. 
\\
When using DeepMIL on top of extracted ImageNet feature vectors, the baseline performance is low at $0.65\pm0.11$ AUC and $0.26\pm0.14$ F1. Using a SimCLR-pretrained feature extractor lifts this to $0.85\pm0.03$ AUC and $0.36\pm0.23$ F1, similar to the results reported in \cite{saillard2021self-owkin-dehaene}. VarMIL further increases this to $0.86\pm0.02$ AUC and $0.47\pm0.10$ F1, which are both significantly higher ($p<0.05$) than SimCLR-DeepMIL, but still significantly lack behind ($p<0.05$) to tile supervision initialized with a SimCLR pretrained feature extractor, especially on the F1 score. SimCLR tile-supervision, however, still lacks behind the top results from \cite{saillard2021self-owkin-dehaene} and \cite{bilal2021development-msi-ranking}.
\\
The ROC and PR curves in Figure \ref{results-roc-crck-msi} and Figure \ref{results-pr-crck-msi} show that all methods that use a SimCLR-pretrained feature extractor far outperform their ImageNet versions. The ROC and PR curves of SimCLR-VarMIL and SimCLR tile supervision can barely be distinguished, however.
\\
Investigation of tiles receiving high and low attention by SimCLR-VarMIL by a pathologist (HMH) revealed that the model pays attention to tiles containing features that are predictive of MSI that were described before by \cite{greenson2009pathologic-predictors-msi}. These results are detailed in the Supplementary material.}

\subsection{Experiment 2: Self-supervised learning and VarMIL improves HRD Score classification in TCGA-BC compared to tile supervision} \label{section-experiment-tcga-bc}
\begin{figure*}[!t]
    \begin{subfigure}[b]{0.33\textwidth}
    \centering
      \includegraphics[width=\linewidth]{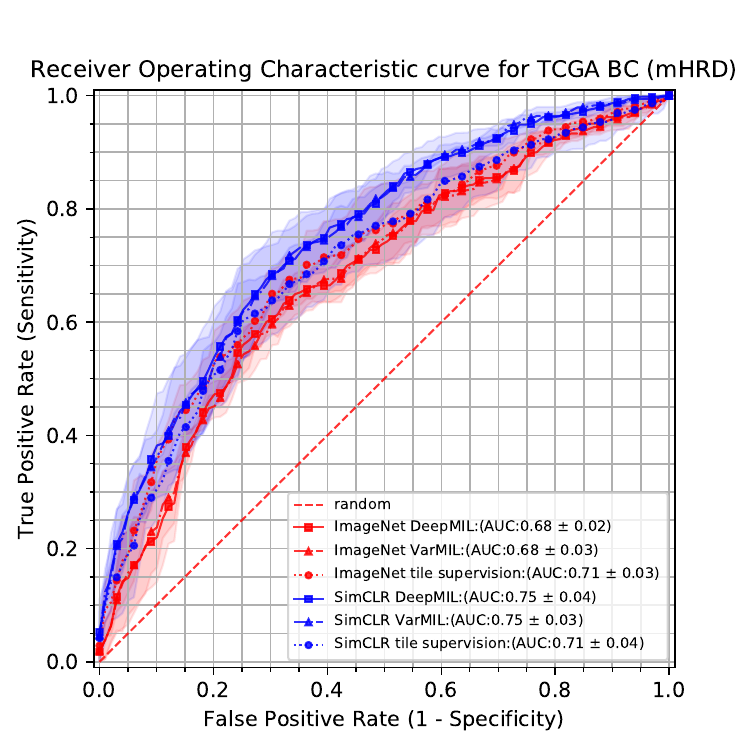}
      \caption{}
      \label{results-roc-bc-mhrd}
    \end{subfigure}
    \hfill
    \begin{subfigure}[b]{0.33\textwidth}
    \centering
      \includegraphics[width=\linewidth]{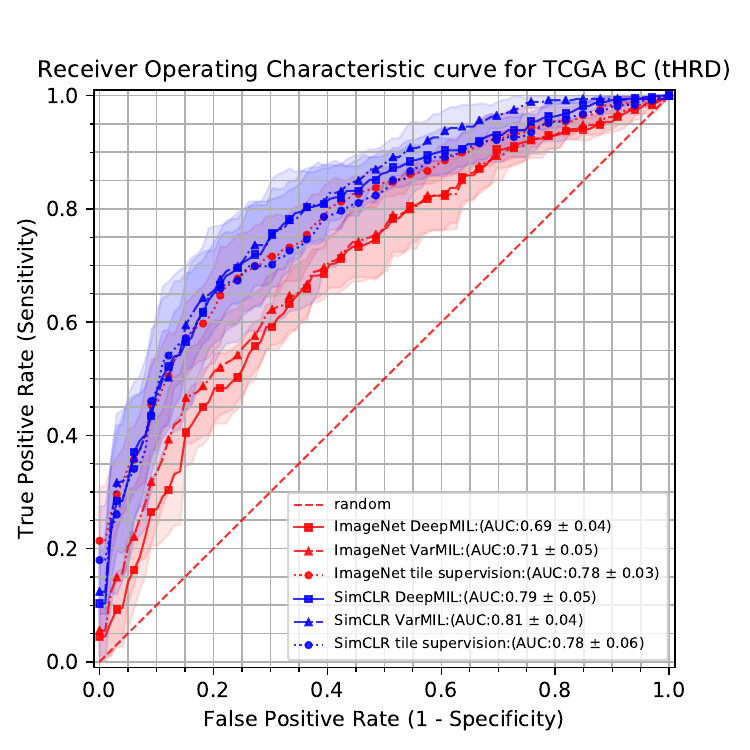}
    \caption{}
    \label{results-roc-bc-thrd}
    \end{subfigure}
    \hfill
    \begin{subfigure}[b]{0.33\textwidth}
    \centering
      \includegraphics[width=1\linewidth]{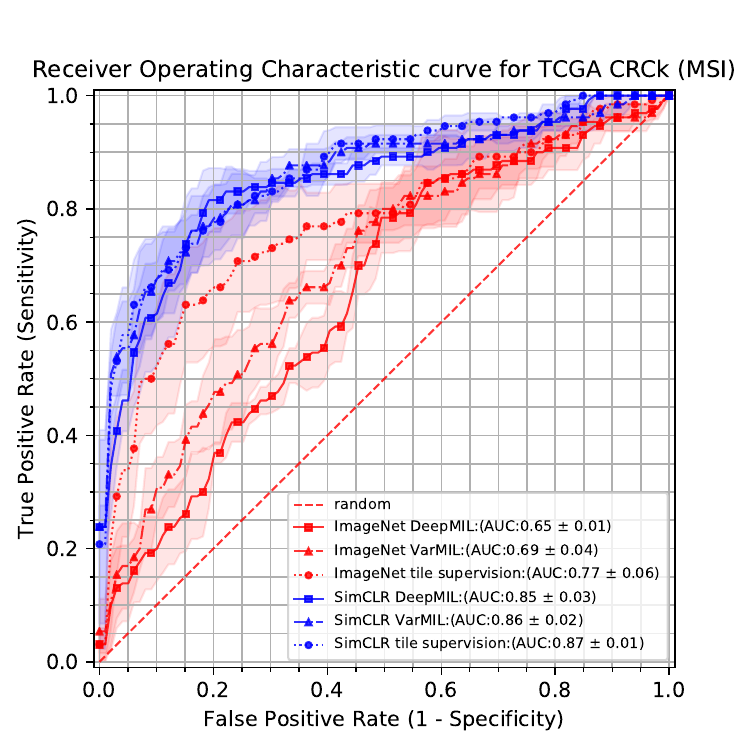}
    \caption{}
    \label{results-roc-crck-msi}
    \end{subfigure}
    \begin{subfigure}[b]{0.33\textwidth}
    \centering
      \includegraphics[width=\linewidth]{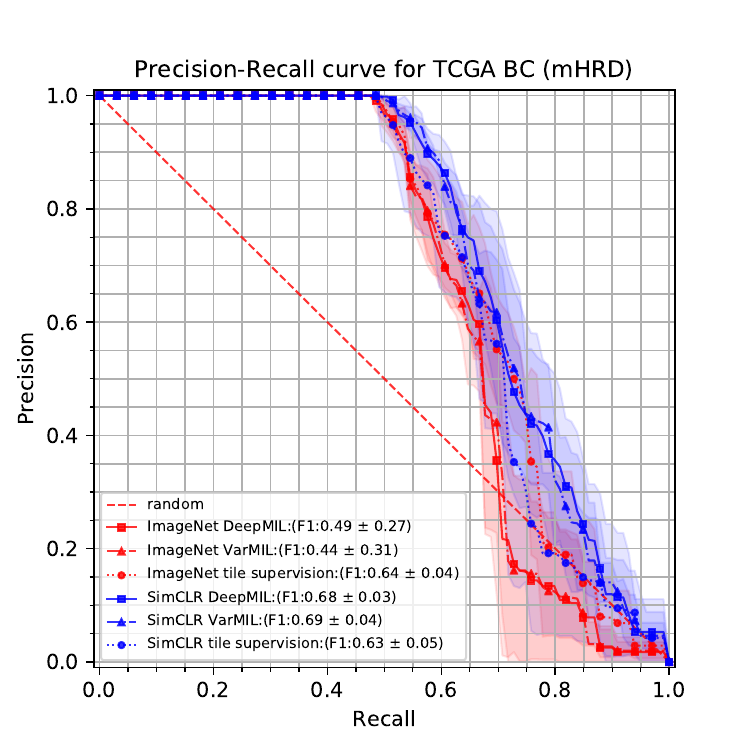}
    \caption{}
    \label{results-pr-bc-mhrd}
    \end{subfigure}
    \hfill
    \begin{subfigure}[b]{0.33\textwidth}
    \centering
      \includegraphics[width=\linewidth]{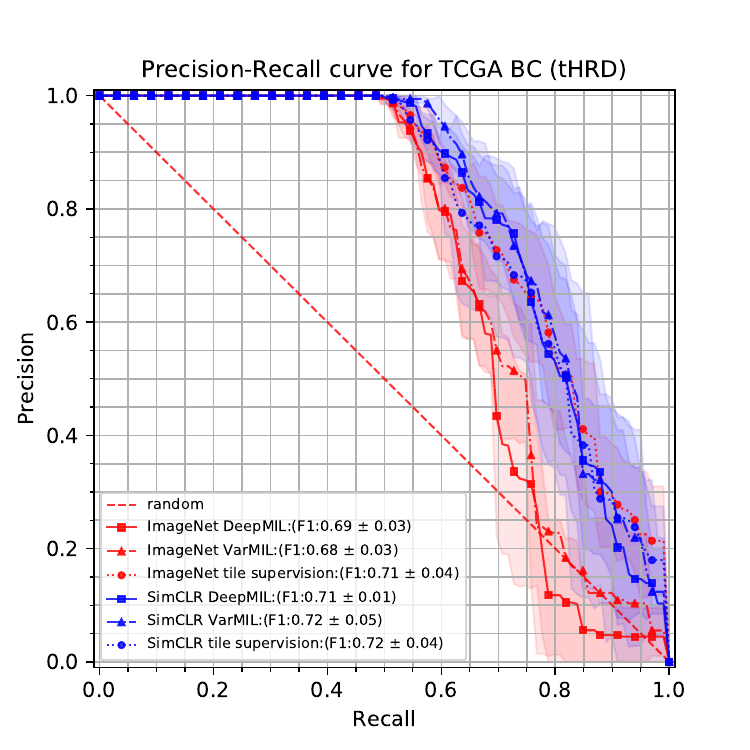}
    \caption{}
    \label{results-pr-bc-thrd}
    \end{subfigure}
    \hfill
    \begin{subfigure}[b]{0.33\textwidth}
    \centering
       \includegraphics[width=1\linewidth]{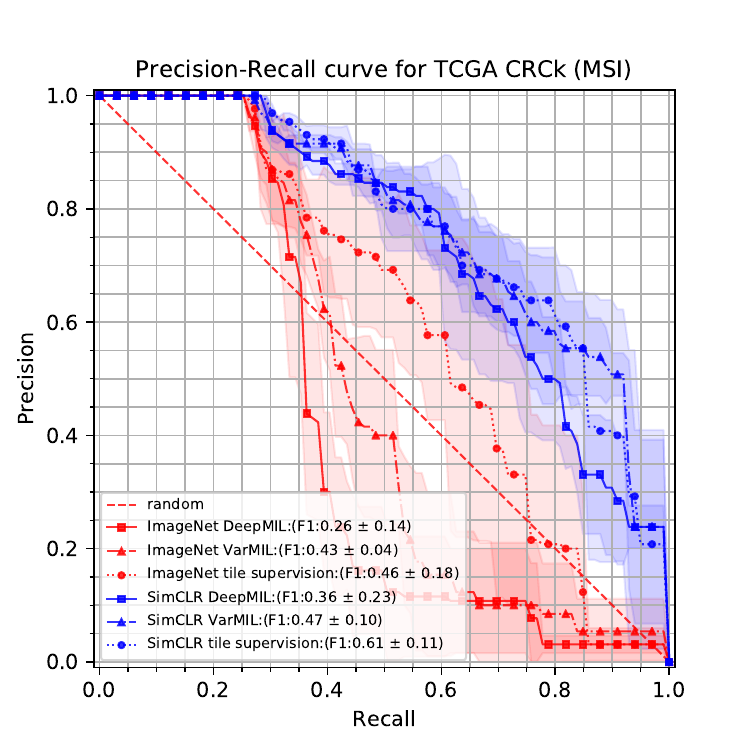}
    \caption{}
    \label{results-pr-crck-msi}
    \end{subfigure}
    \caption{Top row, from left to right: Receiver operating characterstic curves for a) TCGA BC (mHRD), b) TCGA BC (tHRD) and c) TCGA CRCk (MSI) for each combination of feature extractor and classifier method. Bottom row, from left to right: Receiver operating characterstic curves for d) TCGA BC (mHRD), e) TCGA BC (tHRD) and f) TCGA CRCk (MSI) for each combination of feature extractor and classifier method.}
\end{figure*}

\red{
In contrast to Experiment 1, a SimCLR-pretrained feature extractor does not improve performance for tile supervision on the TCGA-BC dataset (for dataset details, see section~\ref{data-tcgabc-basis}) for either binarized HRD labels. As seen in Table \ref{results-table-auc}, tile supervision using either an ImageNet or SimCLR-pretrained feature extractor reaches similar performance at $0.71$ AUC and $0.64$ F1 for mHRD ($p>0.05$ for both) and $0.78$ AUC ($p>0.05$) and $0.71$ to $0.72$ ($p<0.05$) F1 for tHRD. Similarly to Experiment 1, we see that doing WSI compression and MIL using an ImageNet pre-trained extractor generally performs worse than tile supervision, reaching only $0.68\pm0.02$ AUC and $0.46\pm0.27$ F1 for mHRD, and $0.69\pm0.04$ AUC and $0.69\pm0.03$ F1 for tHRD when using DeepMIL. A SimCLR-pretrained extractor and VarMIL boosts this performance, though, up to $0.75\pm0.03$ AUC and $0.69\pm0.04$ F1 for mHRD, and up to $0.81\pm0.04$ AUC and $0.72\pm0.05$ F1 for tHRD. Bootstrapping only reveals a significant difference between the AUC scores of SimCLR-VarMIL and SimCLR-DeepMIL for mHRD, however ($p<0.05$).
\\
Although the differences in the ROC and PR curves are less marked than for MSI, the general observation in Figure \ref{results-pr-bc-mhrd}, \ref{results-pr-bc-thrd}, \ref{results-roc-bc-mhrd}, and \ref{results-roc-bc-thrd} are that the SimCLR-pretrained MIL methods are better across the entire curve than ImageNet MIL methods or any tile supervision method. 
\\
Investigation of tiles receiving high and low attention by SimCLR-VarMIL by a pathologist (HMH) revealed that the model gives high attention to features that are expected to be indicative of a DDR deficiency. These results are detailed in the Supplementary material.
}

\subsection{Experiment 3: Self-supervised learning reduces label requirement by over 50\% }
\begin{figure*}[!t]
    \begin{subfigure}[b]{0.33\textwidth}
    \centering
      \includegraphics[width=\textwidth]{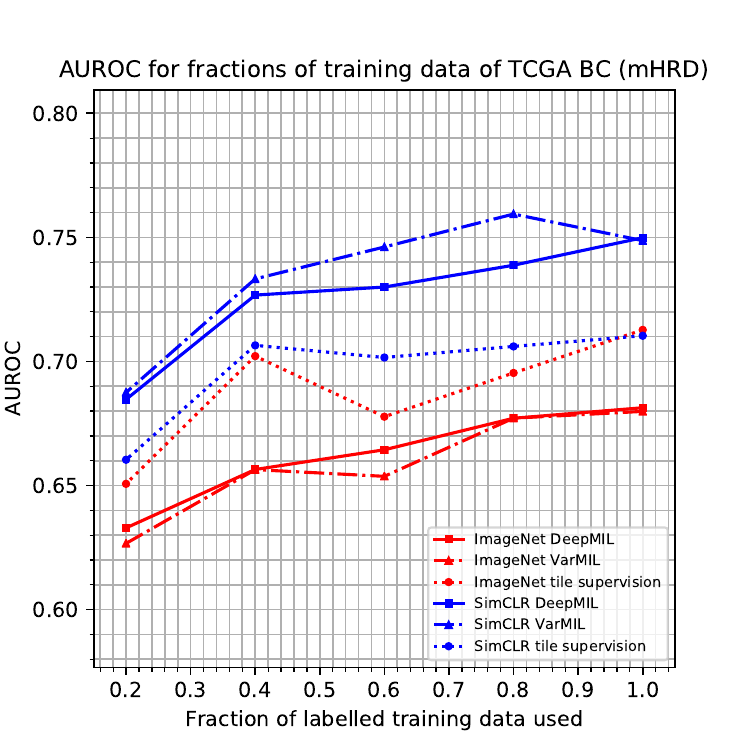}
      \caption{}
      \label{results-fractions-bc-mhrd-auc}
    \end{subfigure}
    \hfill
    \begin{subfigure}[b]{0.33\textwidth}
    \centering
      \includegraphics[width=\textwidth]{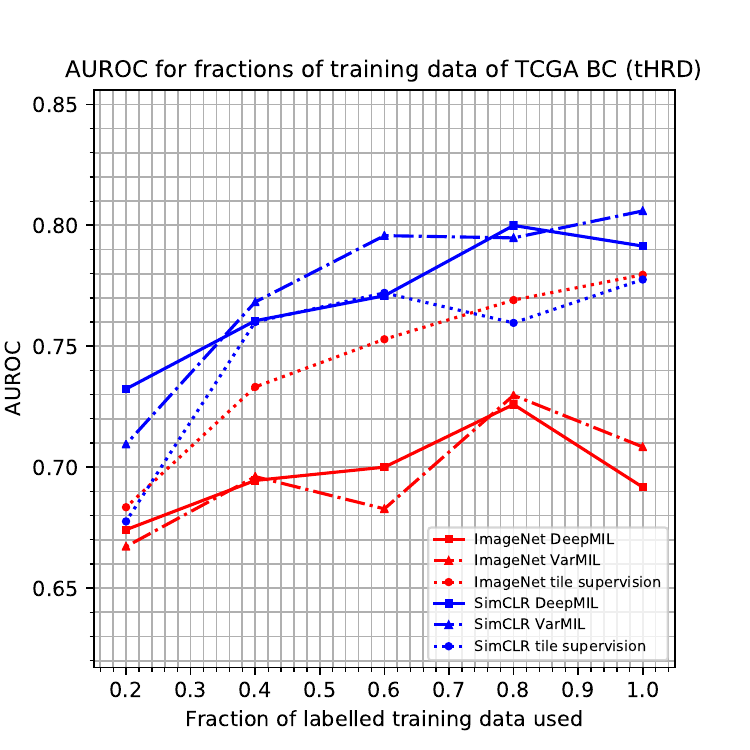}
      \caption{}
      \label{results-fractions-bc-thrd-auc}
    \end{subfigure}
    \hfill
    \begin{subfigure}[b]{0.33\textwidth}
    \centering
      \includegraphics[width=\textwidth]{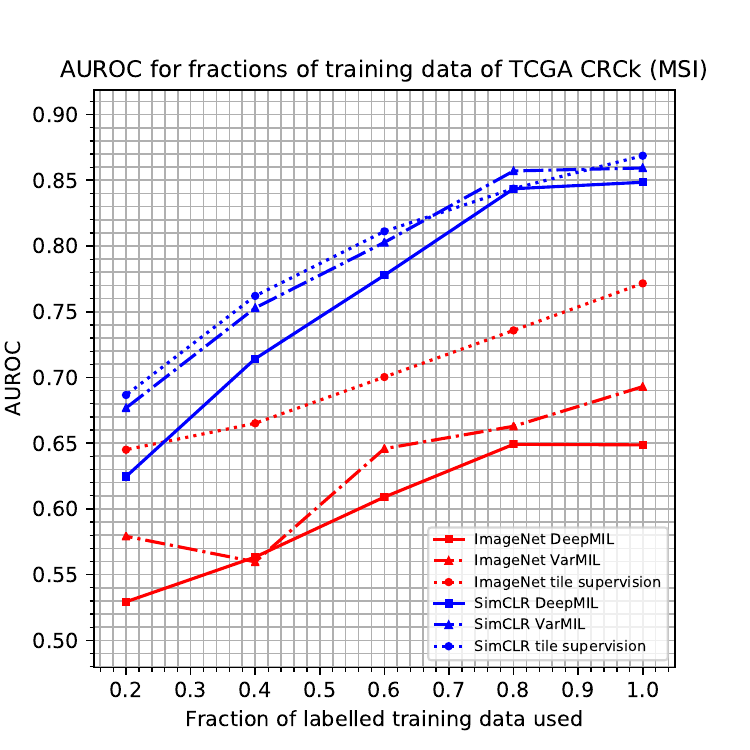}
      \caption{}
      \label{results-fractions-crck-msi-auc}
    \end{subfigure}
    \begin{subfigure}[b]{0.33\textwidth}
    \centering
      \includegraphics[width=1\linewidth]{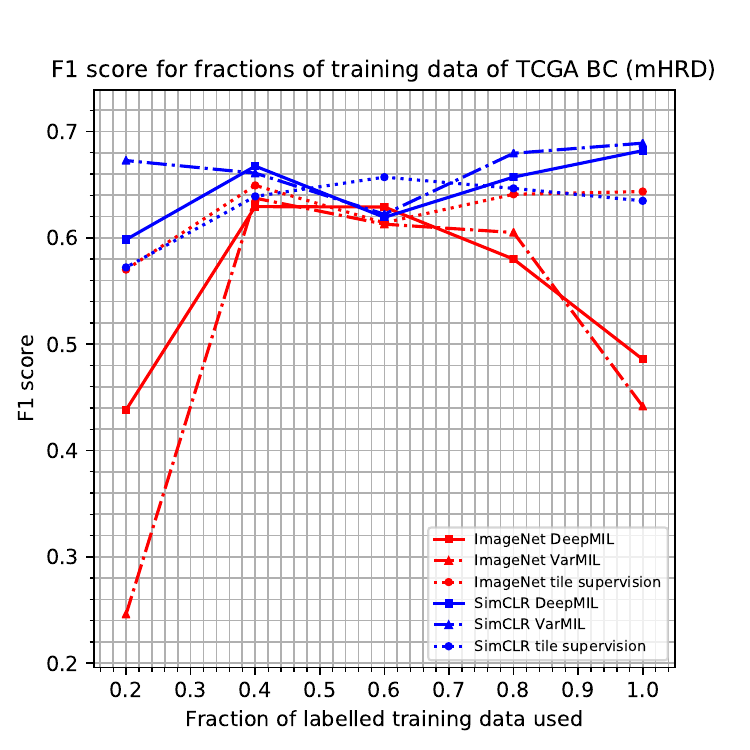}
      \caption{}
      \label{results-fractions-bc-mhrd-f1}
    \end{subfigure}
    \hfill
    \begin{subfigure}[b]{0.33\textwidth}
    \centering
      \includegraphics[width=1\linewidth]{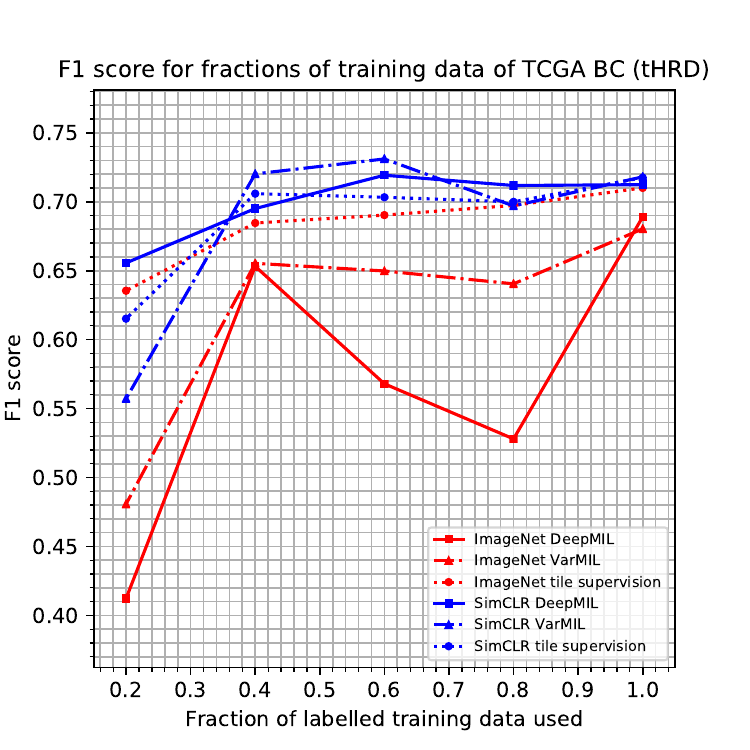}
      \caption{}
      \label{results-fractions-bc-thrd-f1}
    \end{subfigure}
    \hfill
    \begin{subfigure}[b]{0.33\textwidth}
    \centering
      \includegraphics[width=1\linewidth]{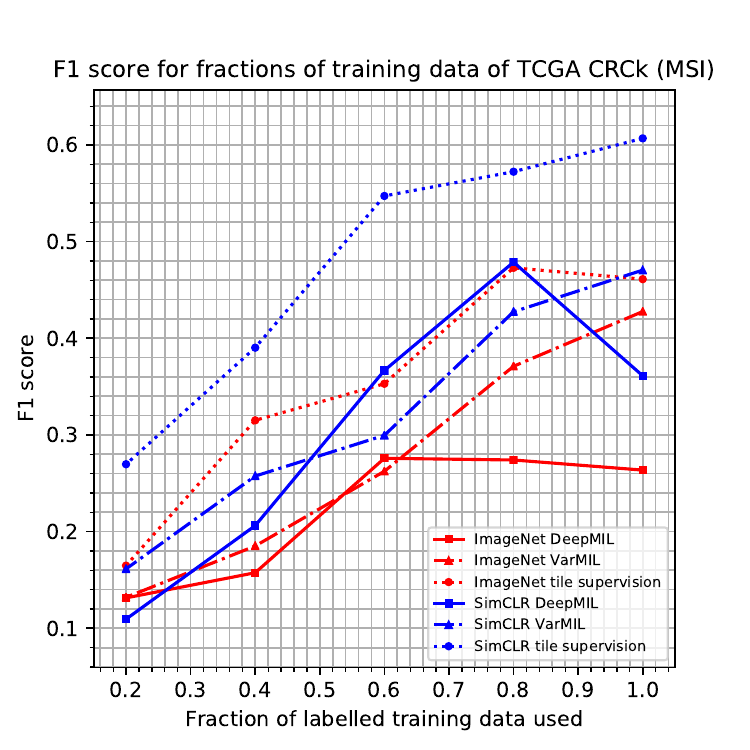}
      \caption{}
      \label{results-fractions-crck-msi-f1}
    \end{subfigure}
    \caption{Top row (AUROC), bottom row (F1), from left to right: achieved metric on test set when trained on $20\%$, $40\%$, $60\%$, $80\%$, and $100\%$ of available labelled training data on (a)/(d) TCGA BC (mHRD), (b)/(e) TCGA BC (tHRD) and (c)/(f) TCGA CRCk (MSI) for each combination of feature extractor and classifier method.}
    \label{results-fractions-all}
\end{figure*}

\red{In Figure \ref{results-fractions-all} we show the AUC (top row, y-axis) and F1 scores (bottom row, y-axis) for every model combination (lines in the graph) of every dataset and label (from left to right: TCGA-BC (mHRD), TCGA-BC (tHRD), TCGA-CRCk (MSI)) when trained on fractions of the labelled training data (x-axis). 
\\
Generally, these results confirm the findings from Experiment 1 and 2, and show that the differences between the models are consistent for varying numbers of training samples. For every method, we see that the AUC improves as more labeled training data is available. The AUC curves for TCGA-BC (mHRD) are relatively flat, likely because the supervisory signal of the median-split HRD labels is relatively weak. For TCGA-BC (tHRD) and TCGA-CRCk (MSI), none of the methods reach a clear point of exhaustion; it seems that all methods would similarly benefit from more data, and the results shown in Experiment 1 and 2 might extrapolate to models trained on larger datasets.
\\
Finally, we see that when using DeepSMILE (SimCLR-VarMIL) on only 40\% of the labeled data, we achieve a higher mean AUC than ImageNet tile-supervision trained on 100\% of the labeled data.
}

\section{Discussion and conclusions} \label{discussion}

We proposed DeepSMILE, which uses a histopathology-specific self-supervised pre-trained feature extractor with VarMIL, a classification network that correctly deals with the weak label and learns an aggregation function over the tiles while modeling intratumor heterogeneity. \red{We have evaluated the added value of a self-supervised pre-trained feature extractor in both the MIL setting where the extractor is frozen and in the tile supervision setting where the extractor is fine-tuned.}

\red{Self-supervised pre-training with SimCLR was shown to be effective in the histopathology domain, in the MIL pipeline on raw TCGA-BC data for HRD prediction, and for both the MIL pipeline and tile supervision pipeline on pre-processed TCGA-CRCk data for MSI prediction. It was valuable even when using the transformations and hyperparameters that were optimized for the ImageNet dataset. Even though pre-training demands more compute time than a single tile supervised training run, training a MIL classifier with frozen features on any downstream task finishes in mere minutes, compared to several hours for tile supervision on a large dataset. For the experiments of this paper, more GPU days were used for the tile-supervision experiments than for the self-supervised pre-training. Additionally we believe that similarly to the pre-trained ImageNet models that are available, the future will see published and shared pre-trained models as done by \cite{ciga2021self-ciga-ssl}. For this purpose, our models can be found online at \url{https://github.com/NKI-AI/hissl} in the model zoo}. 

Only when the tile latent features were extracted with a domain-specific SimCLR pre-trained feature extractor we notice an improved performance with DeepMIL and VarMIL. This performance improvement was not seen when using DeepMIL or VarMIL on top of tile latent feature vectors extracted with an ImageNet pre-trained feature extractor. The tile-supervision baseline, however, \red{only benefited from a histopathology-specific feature extractor for MSI classification on TCGA-CRCk, and not for HRD classification on TCGA-BC. We believe that this is due to the tumor annotations done in TCGA-CRCk, strengthening the supervisory signal for the MSI classification task as most tiles contain tumor tissue that reflects information about the cancer genotype.}

More specifically, when using a SimCLR pre-trained feature extractor, the attention-weighted mean of tile feature vectors as WSI-level latent representation is expressive for HRD and MSI classification. For HRD classification on TCGA-BC, DeepMIL increased AUC by up to $0.04$, and F1 score by up to $0.05$ compared to tile-supervised WSI-label learning. For MSI classification in TCGA-CRCk, self-supervised initialized tile-supervision increased performance by $0.10$ AUC and $0.15$ F1 compared to the ImageNet-initialized tile supervision, outperforming the use of a SimCLR-pretrained feature extractor with DeepMIL.

Finally, an attention-weighted variance of tile feature vectors adds valuable information for HRD and MSI classification. Compared to SimCLR DeepMIL, SimCLR VarMIL increased performance by $0.02$ AUC and $0.01$ F1 for TCGA-BC (tHRD), and by $0.01$ AUC and $0.11$ F1 on TCGA-CRCk (MSI).

\red{For the MSI classification task, our SimCLR-DeepMIL model reproduces the results achieved by \cite{saillard2021self-owkin-dehaene} when using DeepMIL, showing that the use of MOCOv2 or SimCLR is likely equivalent in the histopathology domain. We should note that our SimCLR tile-supervision and VarMIL result is slightly lower than those achieved with Chowder ($0.92$ AUC), although this improved performance doesn't transfer to the full TCGA-CRC dataset. Similarly, \cite{bilal2021development-msi-ranking} achieve a slightly higher score ($0.9$ AUC) on the same dataset \bluetwo{using an ImageNet-initialized extractor and an end-to-end finetuning pipeline.}}

\red{On TCGA-BC, DeepSMILE achieved similar performance on mHRD classification on TCGA-BC as \cite{kather-nature-pan-cancer-histogenomics}, while we did not use tumor bed annotations, stain normalization, quality assessment of the WSIs, and use lower-resolution tiles. We also show that the supervisory signal can be strengthened by setting different cut-off points, which leads to a performance that is similar to that presented in \cite{valieris2020deep-mmrd-hrd}, although a perfect comparison is not possible since they use a different HRD label.}

\red{The relatively high performance with very few available labeled samples indicates that DeepSMILE may be especially useful for tasks where labels are limited, which is the case when one would want to predict, e.g., immune therapy response directly from H\&E WSIs.}

\redtwo{
Finally, the results presented in this work point towards the following recommendations in relation to other work in the literature.
1) We recommend the use of SSL-pretrained feature extractors instead of an ImageNet feature extractor for every WSI-level DDRd classification pipeline in which the tiles are encoded by a frozen feature extractor (e.g. \cite{lu-data-efficient-weakly-supervised-histo}), supported by our results that using DeepMIL on a SimCLR-pretrained feature extractor improved mHRD (AUROC $+7\%$, F1 $+19\%$), tHRD (AUROC $+10\%$, F1 $+2\%$), and MSI (AUROC $+20\%$, F1 $+10\%$) compared to using an ImageNet-pretrained feature extractor. 
2) For tile supervision finetuning pipelines the added value of an SSL-pretrained feature extractor appears to be determined by the strength of the supervisory signal and not the size of the dataset. The TCGA-CRCk dataset with tumor bed annotations benefits from SSL pre-training for tile-supervision (AUROC $+10\%$, F1 $+15\%$), whereas on the TCGA-BC dataset with no tumor bed annotations the noisy supervisory signal appears to impair the benefit of SSL-pretrained initialization (AUROC $+0\%$, F1 $+0\%$ for mHRD). These dynamics are not affected by labelled training dataset size (see Figure \ref{results-fractions-all}).
3) We recommend to use SimCLR or MoCoV2 as a self-supervised learning method, supported by the evidence that our results with SimCLR are similar to the results reported by \cite{saillard2021self-owkin-dehaene} which used MoCoV2 and DeepMIL on the same dataset, whereas the gains with BYOL by \cite{abbasi2021molecular-byol-he} are less apparent.
4) For end-to-end WSI-level label learning pipelines (e.g. \cite{bilal2021development-msi-ranking} we recommend using an SSL-pretrained extractor, since the supervisory signal is strong (only tumor tiles, similar to the annotated TCGA-CRCk dataset combined with WSI-level supervision made possibly by tile subsampling), yet there are only few labelled instances (WSIs) for a multi-million parameter model that is being finetuned. We expect this will further increase the performance of their model further surpassing DeepSMILE's performance.
5) Compared to the tile-supervision pipeline we recommend the DeepSMILE (SimCLR-VarMIL) pipeline as this showed improvements (AUROC: $+4\%$ to $+9\%$, F1 $+1\%$ to $+5\%$) across both evaluated datasets.
6) Compared to DeepMIL, we would recommend to use VarMIL on problems where it is likely that heterogeneity of tiles may be predictive of the label, as it hints at achieving a slightly higher AUROC and F1 ($0.01$ to $0.02$ across datasets, $p>0.05$). 
7) We can not necessarily recommend VarMIL over CLAM \citep{lu-data-efficient-weakly-supervised-histo} since the addition of the cluster constraint leads to similar performance increases as the feature variability to DeepMIL, with an increase of $+0.01$ to $+0.02$ AUC with 100\% of labelled data and with an increased benefit in low labelled data regimes for some tasks (e.g. Fig \ref{results-fractions-crck-msi-auc}).
}

\subsection{Limitations}
We acknowledge several limitations to the experimental setup and our proposed method.

\blue{Firstly, we have not explicitly investigated the added effect of self-supervised learning and VarMIL relative to varying pre-processing steps like tumor annotation or H\&E normalization. We expect self-supervised pre-training to reduce the positive effect of stain normalization, and VarMIL to reduce the positive effect of tumor bed annotations. With our current datasets, we can not confirm the hypothesis that tile-supervision benefits from a SimCLR-pretrained network in TCGA-CRCk but not in TCGA-BC due to the tumor annotations. This should be further investigated in future work.}

\red{Similarly, this work has not investigated the improved generalizability or robustness of models that use a SimCLR-pretrained feature extractor when evaluating the models on external datasets. We expect that these feature extractors are less prone to shifts in H\&E color distribution or microscope imaging features due to their heavy data augmentations and inability to overfit on any specific label, but this should be investigated in future work.}

Thirdly, the performance of our proposed method can likely be improved by using wider and deeper feature extractors, pre-training for longer on a larger set of pan-cancer WSIs, and with domain-optimized hyperparameters and data augmentations for self-supervised pre-training in the histopathology domain. Similarly, in our current set-up, the feature extractor is not further fine-tuned for the task at hand when using DeepMIL or VarMIL. 

\red{Furthermore, one should remain aware that these models do not necessarily find causal relationships between morphological features and HRD or MSI. Although we find that DeepSMILE focuses on tiles that contain features that are likely to be related to the genomic markers (see Supplementary material), cancer is a genetic disease which, through a complicated interplay of gene expression, antigen production, and immune system response, leads to complex cellular and tumor microenvironment phenotypes. Likely, features like immune infiltrate, necrosis, and tumor heterogeneity are detected by the model which are correlated with DDRd labels, yet these features might also be present in tumors not having these specific genetic features. This need not be a problem for the grander goal of, e.g., immune therapy response prediction, however, since such morphological features in the tumor microenvironment are increasingly found to be an important player when predicting therapy response \citep{blank2016cancer}. For example, \cite{mlecnik2016integrative-crc-msi-immuno} show that MSS (MSI) patients with high (low) immune infiltrate will (not) benefit from immune therapy.}

Finally, although VarMIL models inter-tile feature heterogeneity, the model does not take into account feature interactions, multiple resolutions, or the spatial context of the tiles. Future research could investigate and compare the effectiveness and computational requirements of Neural Image Compression (\cite{tellez-nic-gigapixel}) and end-to-end context-aware learning (\cite{pinckaers-streaming-original, chen-full-wsi-learning}) for genomic label classification to our proposed method. Additionally, it could be interesting to investigate the performance of vision transformers \citep{dosovitskiy2020image-vit-vision-transformers} on top of extracted latent feature vectors of all tiles of a WSI to model tile interactions with low computational complexity.

\subsection{Conclusions} \label{conclusions}
We have shown that self-supervised learning and weak label learning methods in computational histopathology can lift the performance of HRD and MSI classification directly from H\&E WSIs \red{on similar sized datasets, or reach the same performance as \bluetwo{the ImageNet-initialized tile supervision} baseline with only 40\% of the labeled data}.

In the future, these methodological improvements may reduce the need for expensive genome sequencing techniques, provide personalized therapy recommendations based on widely available H\&E WSIs, and improve patient care with quicker treatment decisions - also in medical centers without access to genome sequencing resources.

\section*{Acknowledgements}
The collaboration project is co-funded by the PPP Allowance made available by Health~Holland\footnote{\url{https://www.health-holland.com}}, Top Sector Life Sciences \& Health, to stimulate public-private partnerships.

\section*{Conflicts of interest}
The authors declare the following competing interests. E.G. is a shareholder of Ellogon.ai. H.M. Horlings received a consultation fee paid to the institute from Roche Diagnostics, outside the scope of the work described here. The other authors report no conflicts of interest.

\bibliographystyle{model2-names.bst}\biboptions{authoryear}
\bibliography{ms}

\clearpage


\section*{Supplementary material}

**Supplementary material for "DeepSMILE: Contrastive self-supervised pre-training benefits MSI and HRD classification directly from H&E whole-slide images in colorectal and breast cancer"**

**Visualization of tiles indicative of the label of interest.**

We have used the predictions of SimCLR-VarMIL of the first fold on the test for both TCGA-BC and TCGA-CRCk. We then visualize once without cherry-picking:
- for positive and negative labels
- of the top 6 slides (highest WSI-level prediction score for positive labels, lowest WSI-level prediction for negative labels
- the 1st and 5th highest and lowest attention tiles

These images can be viewed on the next four pages of this Supplementary material.

A pathologist (HMH) has viewed these images, compared the high- to low attention tiles, and compared the high attention tiles of positive labels to the high attention tiles of negative labels. In summary, we find similar features as were shown before for MSI (Greenson et al (2009)**),** and find features in HRD patients that are expected for HRD patients. A detailed description of the findings is as follows:

For MSI-H and MSS, we find that the visualized tiles that received high attention all contain colorectal tumorous tissue, whereas most of the low attention tiles contain normal epithelium and fibro-adipose connective tissue**.** For MSI-H, we see that these tumour-containing tiles also display a high level of necrosis and mucinous differentiation. These are features described ealier by Greenson et al (2009) and also shown to be present in the models by, e.g., Saillard et al (2021). Surprisingly, however a high number of tiles with low attention contain immune infiltrate, whereas we expected these to be given high attention.

For HRD, we find again that, for both HRD-low and HRD-high patient, high attention is given to tiles containing tumor tissue and low attention is given to tiles containing blood vessels and connective tissue consisting of collagen and fat cells. The high attention tiles in HRD-high patients, however, mostly contain tumor cells that show high nuclear pleomorphism, whereas the HRD-low patients contain tiles that appear to be mucinous or invasive lobular carcinoma. Additionally, the high attention tiles of HRD-high patients, compared to those of HRD-low patient, contain unorganized and heterogeneous tumor cells, many extravasated blood cells probably due to necrosis.

We have added these visualizations as supplementary material, and added a paragraph in the results/discussion. We are aware, however, that the results might be biased due to tile selection and only having a single pathologist look at these tiles. In future research, we think that it would be a good idea to perform a blinded experiment with several pathologists to investigate in more detail the association between the morphological features found in the tiles that the model pays attention to and these genomic labels.

TCGA-CRCk
MSI-h cases with high prediction
High attention • Low attention

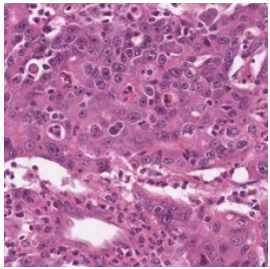 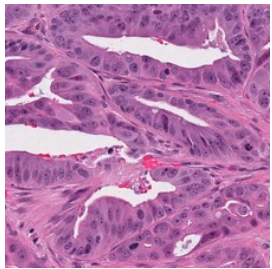 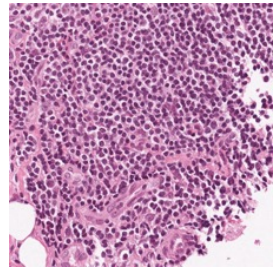 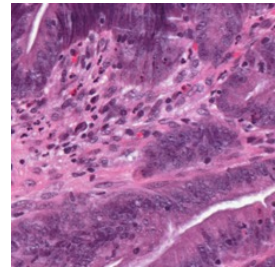
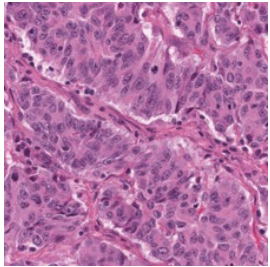 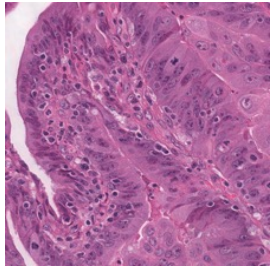 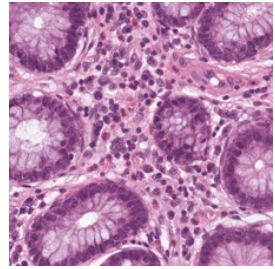 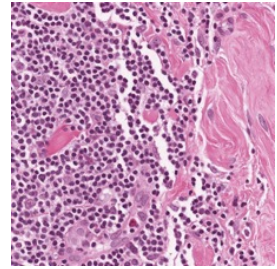
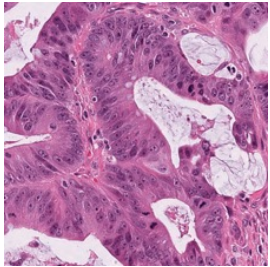 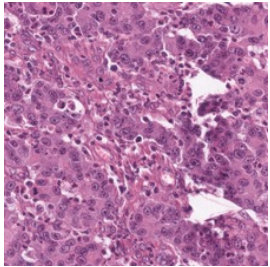 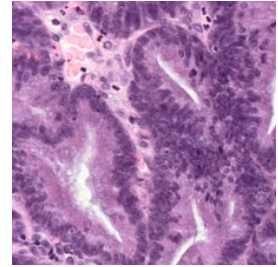 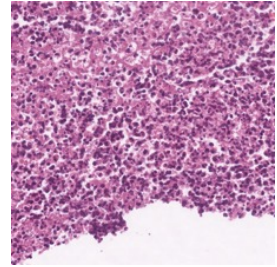
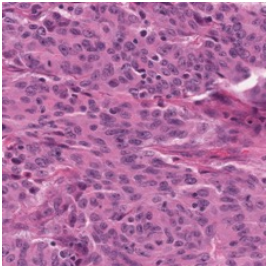 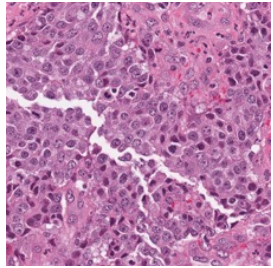 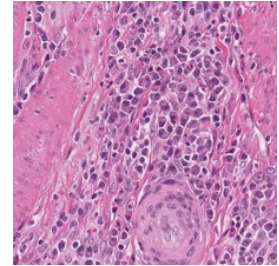 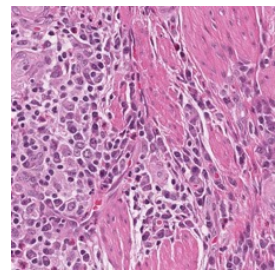
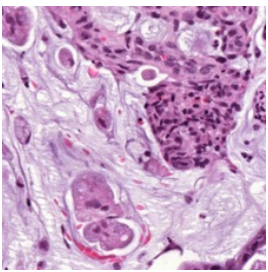 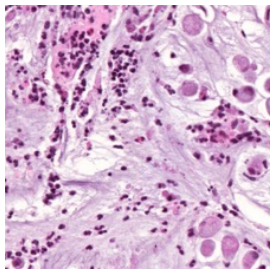 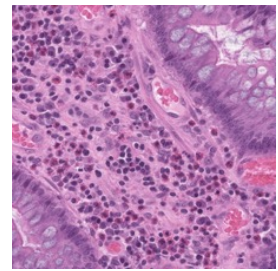 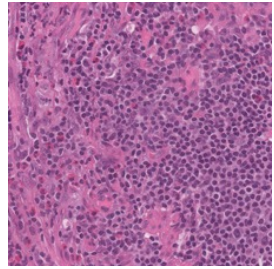
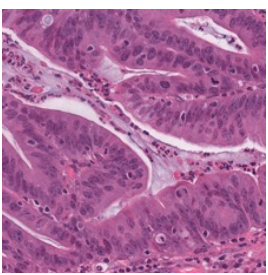 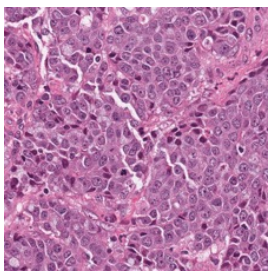 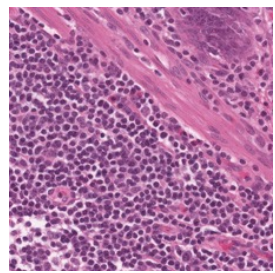 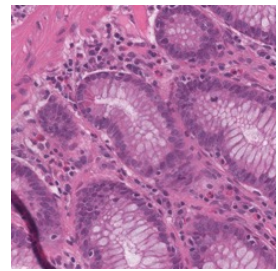

## TCGA-CRCk
MSS cases with low prediction

<u>High attention</u> <u>Low attention</u>

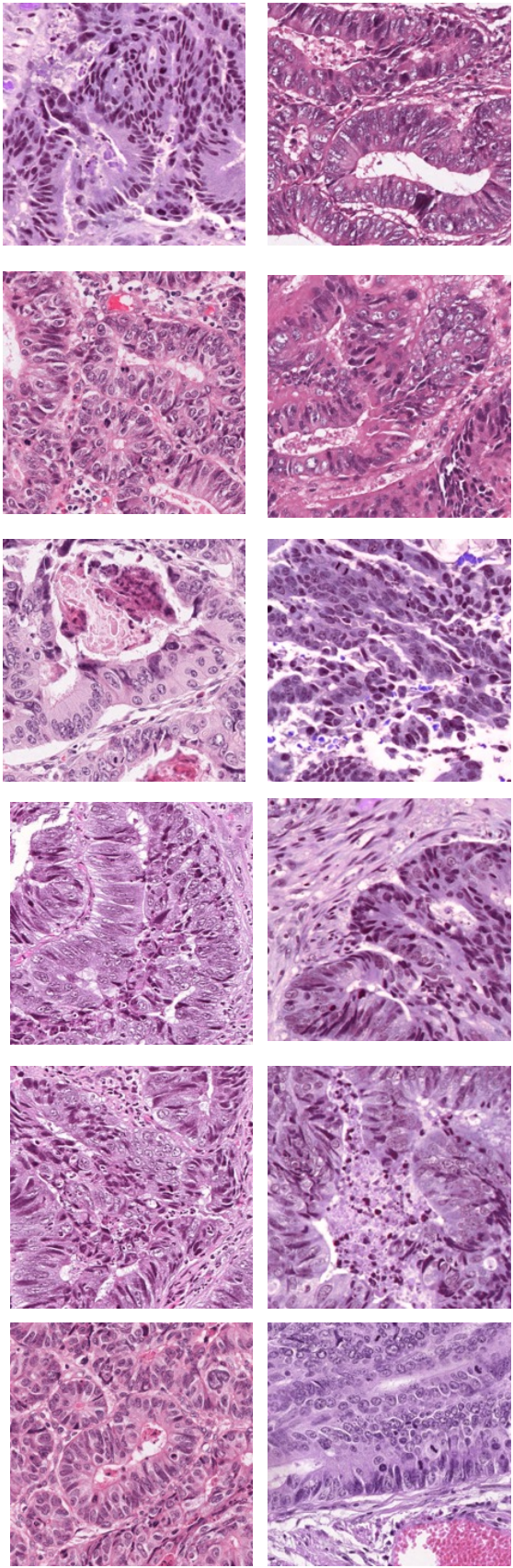 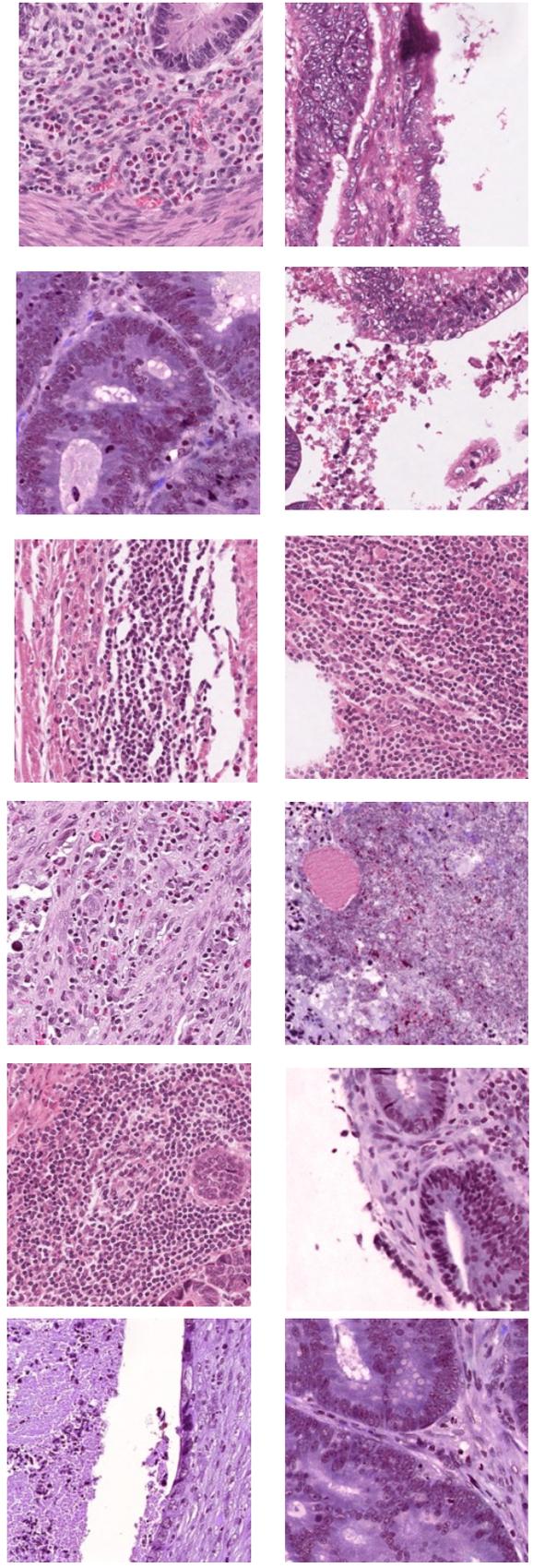

# TCGA-BC
<u>High attention</u>   tHRD=1 cases with high prediction   <u>Low attention</u>

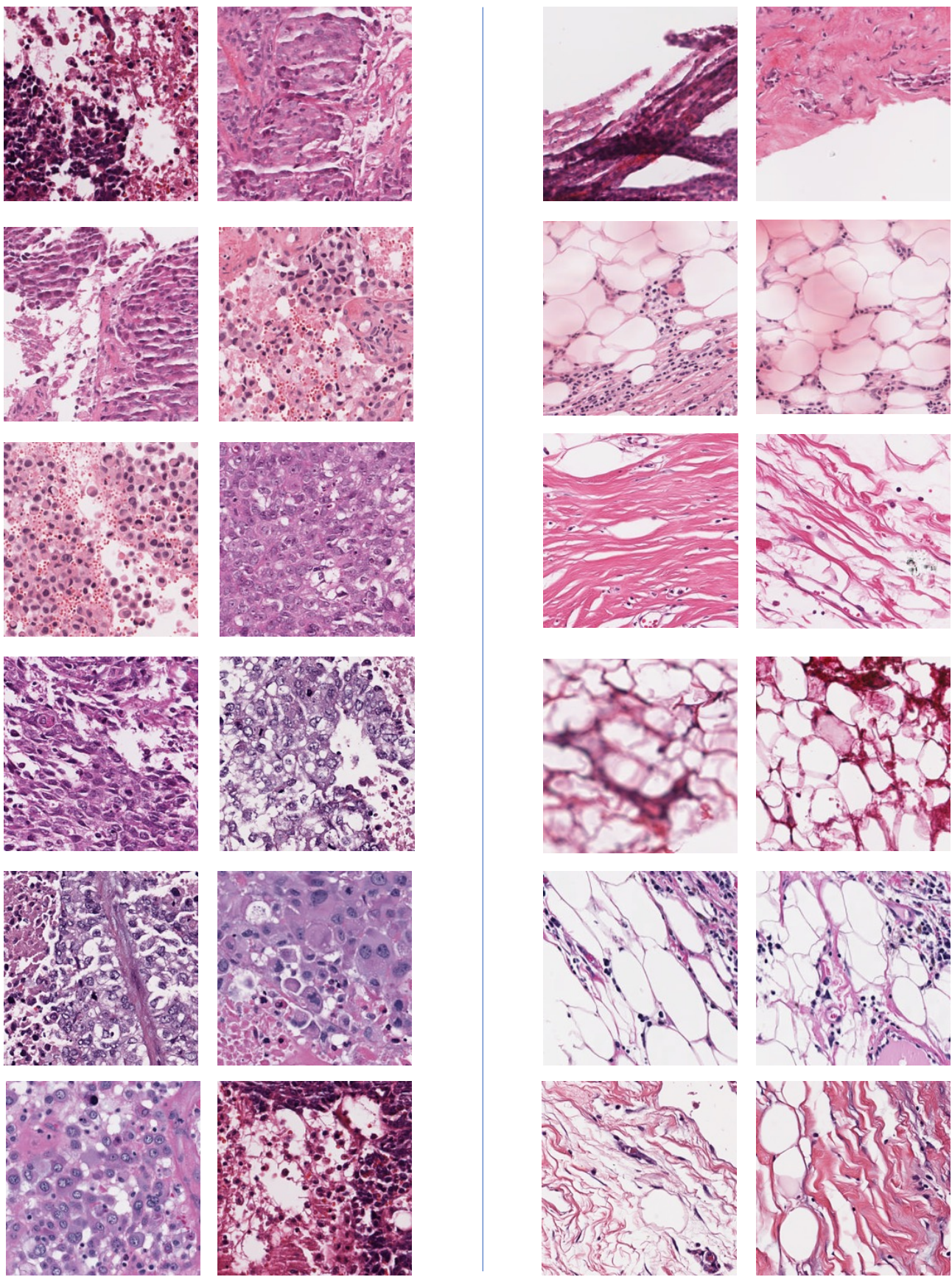

TCGA-BC
High attention    tHRD=0 cases with low prediction    Low attention

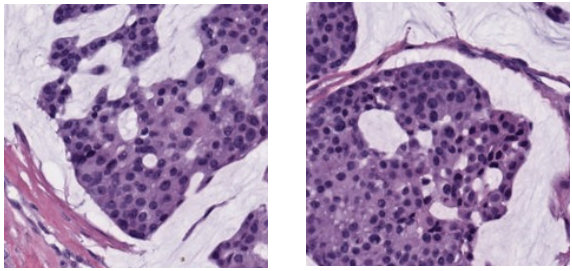
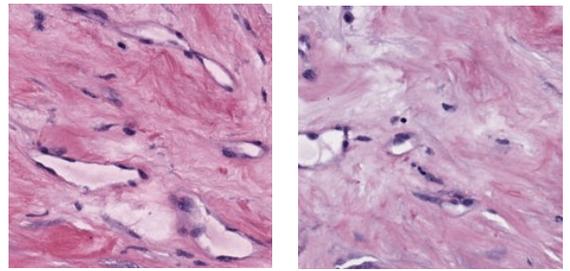
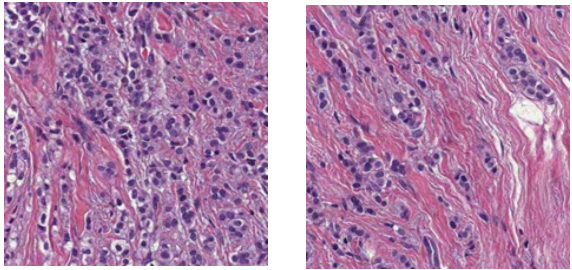
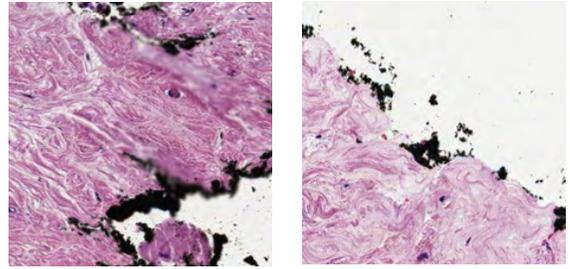
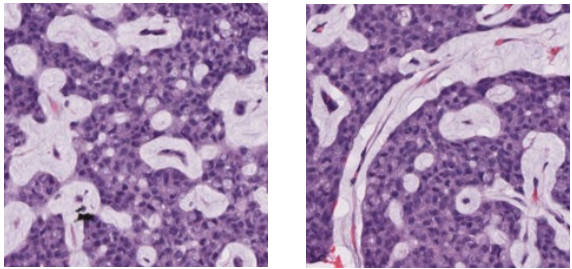
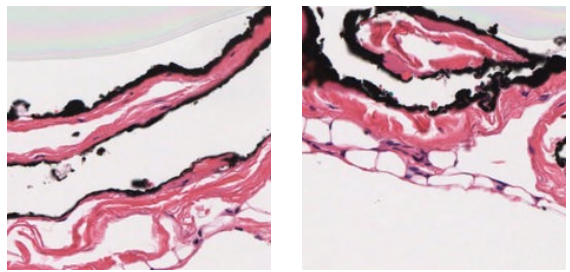
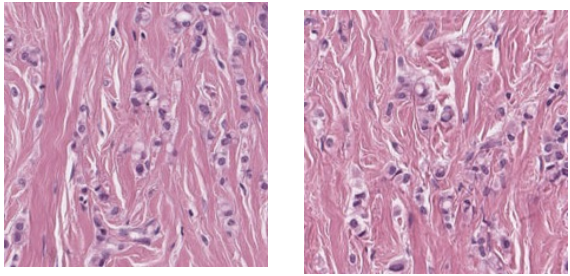
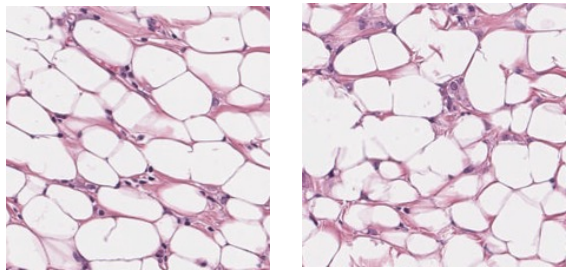
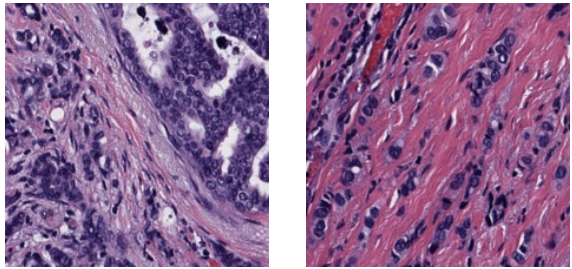
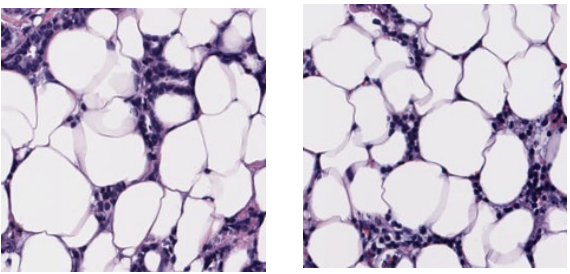
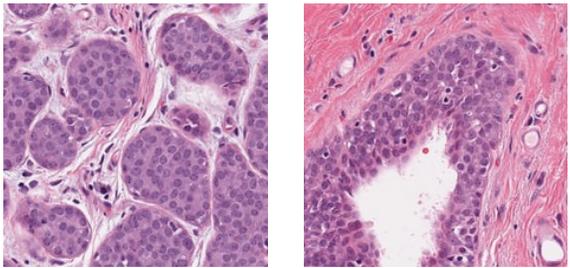
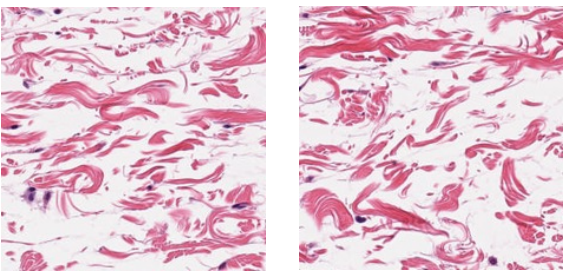


\clearpage

\end{document}